\documentclass[journal]{IEEEtran}

\usepackage{cite} 

\ifCLASSINFOpdf
  \usepackage[pdftex]{graphicx}
  \graphicspath{{images/}}
  \DeclareGraphicsExtensions{.pdf,.jpeg,.png}
\else
  \usepackage[dvips]{graphicx}
  \graphicspath{{images/}}
  \DeclareGraphicsExtensions{.eps}
\fi  
  
\usepackage[cmex10]{amsmath}
\usepackage{amssymb}
\usepackage{amsthm}

\usepackage{algpseudocode} 

\usepackage{array}

\usepackage{mathrsfs}
\DeclareMathAlphabet{\mathpzc}{OT1}{pzc}{m}{it}
\usepackage{bm}

\usepackage[caption=false,font=footnotesize]{subfig}

\usepackage{url}

\usepackage{fancybox}

\usepackage{calc}
\newsavebox\myboxa

\usepackage{overpic}
\usepackage{psfrag}

\usepackage{dblfloatfix}

\usepackage{txfonts}
\usepackage{textcomp}

\usepackage{stmaryrd}

\usepackage{pbox}

\usepackage{diagbox}

\usepackage{xcolor}
\usepackage{colortbl}

\usepackage{hhline}

\usepackage{multirow}

\usepackage{makecell}

\usepackage{pifont}
\newcommand{\cmark}{\ding{51}}%
\newcommand{\xmark}{\ding{55}}%

\usepackage{soul}

\makeatletter
\newsavebox\saved@arstrutbox
\newcommand*{\setarstrut}[1]{%
  \noalign{%
    \begingroup
      \global\setbox\saved@arstrutbox\copy\@arstrutbox
      #1%
      \global\setbox\@arstrutbox\hbox{%
        \vrule \@height\arraystretch\ht\strutbox
               \@depth\arraystretch \dp\strutbox
               \@width\z@
      }%
    \endgroup
  }%
}
\newcommand*{\restorearstrut}{%
  \noalign{%
    \global\setbox\@arstrutbox\copy\saved@arstrutbox
  }%
}
\makeatother

\makeatletter

\newcommand{\Rmnum}[1]{\expandafter\@slowromancap\romannumeral #1@}
\makeatother

\begin{document}

\title{WISERNet: Wider Separate-then-reunion Network for
  Steganalysis of Color Images}

\author{ Jishen Zeng,~\IEEEmembership{Student Member,~IEEE,} 
  Shunquan~Tan*,~\IEEEmembership{Senior Member,~IEEE,}
  Guangqing~Liu, \\
  Bin~Li,~\IEEEmembership{Senior Member,~IEEE,}
  and~Jiwu~Huang,~\IEEEmembership{Fellow,~IEEE}%

  \thanks{J. Zeng and J. Huang are with the Guangdong Key Laboratory of
    Intelligent Information Processing and Shenzhen Key Laboratory of
    Media Security, also National Engineering Laboratory for Big Data
    System Computing Technology, Shenzhen University, Shenzhen 518060,
    China (email: jishenzeng@foxmail.com; jwhuang@szu.edu.cn).}

  \thanks{S. Tan and G. Liu are with College of Computer Science and
    Software Engineering, Shenzhen University. They are also with with
    the Guangdong Key Laboratory of Intelligent Information
    Processing, Shenzhen Key Laboratory of Media Security, and
    National Engineering Laboratory for Big Data System Computing
    Technology, Shenzhen University, Shenzhen 518060, China (email:
    tansq@szu.edu.cn; l1294496804@qq.com).}

  \thanks{B. Li is with Guangdong Key Laboratory of Intelligent
    Information Processing and Shenzhen Key Laboratory of Media
    Security, Shenzhen University, Shenzhen 518060, China, and also
    with Peng Cheng Laboratory, Shenzhen 518052, China (email:
    libin@szu.edu.cn).}

  \thanks{*S.~Tan is the correspondence author.}

  \thanks{This work was supported in part by the NSFC~(61772349,
    U1636202, 61572329, 61872244), Shenzhen R\&D
    Program~(JCYJ20160328144421330). This work was also supported by
    Alibaba Group through Alibaba Innovative Research (AIR) Program.}%
}

\maketitle
\begin{abstract}
  Until recently, deep steganalyzers in spatial domain have been all
  designed for gray-scale images. In this paper, we propose
  WISERNet~(the wider separate-then-reunion network) for steganalysis
  of color images. We provide theoretical rationale to claim that the
  summation in normal convolution is one sort of ``linear collusion
  attack'' which reserves strong correlated patterns while impairs
  uncorrelated noises. Therefore in the bottom convolutional layer
  which aims at suppressing correlated image contents, we adopt
  separate channel-wise convolution without summation
  instead. Conversely, in the upper convolutional layers we believe
  that the summation in normal convolution is beneficial. Therefore we
  adopt united normal convolution in those layers and make them
  remarkably wider to reinforce the effect of ``linear collusion
  attack''. As a result, our proposed wide-and-shallow,
  separate-then-reunion network structure is specifically suitable for
  color image steganalysis. We have conducted extensive experiments on
  color image datasets generated from BOSSBase raw images and another
  large-scale dataset which contains 100,000 raw images, with
  different demosaicking algorithms and down-sampling algorithms. The
  experimental results show that our proposed network outperforms
  other state-of-the-art color image steganalytic models either
  hand-crafted or learned using deep networks in the literature by a
  clear margin. Specifically, it is noted that the detection
  performance gain is achieved with less than half the complexity
  compared to the most advanced deep-learning steganalyzer as far as
  we know, which is scarce in the literature.
\end{abstract}


\bstctlcite{csdl_bstctl}

\section{Introduction}
\label{sec:intro}

\IEEEPARstart{I}{n} the last decade, the main battleground of
spatial image information hiding is at gray-scale cover
images. However, the confrontation between color image steganography
and its rival, color image steganalysis has drawn ever greater
attentions of researchers due to the fact that the vast majority of
digital images in real life have colors. 

Most of the modern gray-scale steganographic algorithms including
famed SUNIWARD~\cite{holub_eurasip_2014}, HILL~\cite{li_icip_2014} and
MiPOD~\cite{sedighi_tifs_2016} adopt the so-called additive embedding
distortion minimizing framework~\cite{fridrich_spie_2007}. Move a step
further, Denemark et al.~\cite{denemark_ihmmsec_2015} and Li et
al.~\cite{li_tifs_2015}~(named CMD steganography) independently
constructed an effective approach to preserve the correlation between
neighboring pixels. Generally, gray-scale image steganographic
algorithms, such as SUNIWARD and HILL can be directly used for color
images, by treating every color band~\footnote{In the literature, the
  red/greeen/blue channels in true-color images are also called
  ``color bands''. Throughout the paper, we adopt the term ``color
  bands'' to avoid confusion with the term ``channels'' in
  deep-learning architectures.} as a gray-scale image and embedding
secret bits into the bands independently. There has been no
steganographic algorithms aiming at color images until recently. In
2016, inspired by CMD steganography, Tang et al. proposed a so-called
CMD-C steganography~\cite{tang_spl_2016} purposely for color images.
CMD-C can preserve not only the correlation within each color band,
but also the correlation among three color bands. Therefore it can
better resist steganalysis targeted for color images. We denote the
CMD-C steganography using SUNIWARD~\cite{holub_eurasip_2014} and
HILL~\cite{li_icip_2014} for initialization as CMD-C-SUNIWARD and
CMD-C-HILL, respectively.~\footnote{Throughout this paper, the
  acronyms used for the steganographic and steganalytic algorithms are
  taken from the original papers. The corresponding full names are
  omitted for brevity.}

The current dominant universal/blind gray-scale steganalytic detectors
use \textit{rich models} with tens of thousands of
features~\cite{fridrich_tifs_2012,holub_tifs_2013}, and an ensemble
classifier~\cite{kodovsky_tifs_2012}. Rich models for color image
steganalysis had been proposed even before specialized color
steganography arose. In 2014, Goljan et al. proposed the first color
rich-model steganalytic features
set~(CRM)~\cite{goljan_wifs_2014}. Since most of the digital color
images are captured by cameras with one single sensor plus a Color
Filter Array~(CFA), the CFA demosaicking algorithm used in the
generation procedure introduces constraints on the relationship
between neighboring color pixels, which can be utilized in
steganalysis. In \cite{goljan_spie_2015}, Goljan et al. proposed a
CFA-aware rich model for color image steganalysis. In
\cite{abdulrahman_iwcc_2015}, Abdulrahman et al. proposed another
CFA-aware rich model. Further on, they fused CRM with features
extracted from inter-band geometric transformation
measures~(GCRM)~\cite{abdulrahman_scn_2016}, and features based on
steerable Gaussian filters
bank~(SGRM)~\cite{abdulrahman_ihmmsec_2016}. However, the
applicability of the features proposed in
\cite{goljan_spie_2015,abdulrahman_iwcc_2015,abdulrahman_scn_2016,abdulrahman_ihmmsec_2016}
remains limited due to the fact that varying post-processing
procedures, e.g. down-sampling and rotating, tend to severely weaken
CFA related correlations.

Characteristics of the specific attacking
targets~\cite{denemark_wifs_2014,tang_tifs_2016} and the specific
cover sources~\cite{boroumand_tifs_2018} can be utilized to further
improve detection performance of rich models. But, utilizing the
knowledge of attacking targets might violate the purpose of
universal/blind steganalysis, while utilizing the characteristics of
specific cover sources might lead to over-optimized
detectors. Therefore in our work we adhere to generic universal
steganalysis with neither the knowledge of specific steganographic
algorithms~(e.g. content-adaptive) nor the knowledge of specific cover
sources.

In recent years, deep-learning networks have achieved overwhelming
superiority over conventional approaches in many
fields~\cite{schmidhuber_nn_2015}. Researchers in image steganalysis
have also tried to investigate the potential of deep-learning networks
in this field. Started from the pioneer work of Tan and
Li~\cite{tan_apsipa_2014}, researchers have kept trying to promote
detection performance of deep-learning
steganalyzers~\cite{qian_spie_2015,pibre_ei_2016_ver2,qian_icip_2016}.
In 2016, Xu et al. proposed a Convolutional Neural Network~(CNN)
structure for image steganalysis which outperforms hand-crafted
steganalytic features~\cite{xu_spl_2016}~(referred as Xu's model \#1).
In the following years, the literature witnessed deep-learning
steganalyzers going deeper and more complicated. Ye et al. proposed a
deeper CNN equipped with a new activation function~(TLU) which can
further boost detection performance~\cite{ye_tifs_2017}~(referred as
Ye's model). In JPEG domain, Zeng et al. proposed a hybrid CNN
steganalyzers equipped with quantization and truncation, which is
obviously superior to hand-crafted JPEG steganalytic
features~\cite{zeng_ei_2017,zeng_tifs_2018}. Chen et al. proposed a
JPEG-phase-aware deep CNN steganalyzer~\cite{chen_ihmmsec_2017}.
Inspired by ResNet~\cite{he_cvpr_2016}, Xu proposed a much deeper CNN
based JPEG steganalyzer~\cite{xu_ihmmsec_2017}~(referred as Xu's model
\#2). However, all of the above deep-learning steganalyzers are
gray-scale image oriented. According to our best knowledge, there is
no report to address deep-learning based color image steganalysis.

In this paper, we propose WISERNet, a specific wider
separate-then-reunion network for steganalysis of color images. We
claim that the summation in normal convolution is one sort of ``linear
collusion attack'' which is the process of forming a linear
combination of input bands~\cite{su_tm_2005}. It reserves strong
correlated patterns while impairs uncorrelated noises. Since the main
purpose of the convolutions in the bottom convolutional layer is to
suppress correlated image contents, for the bottom convolutional layer
we discard summation and introduce channel-wise convolution. On the
other hand, in the upper convolutional layers we still adopt normal
convolution which retains summation and make them remarkably wider, in
order to reinforce the effect of ``linear collusion attack''. This is
because we believe such an upper structure is beneficial to the
ability of CFA related ``hidden-pattern-aware''.  We have conducted
extensive experiments on color image datasets generated from public
true-color raw images as well as a large-scale dataset which contains
100,000 raw images. The results demonstrated the superiority of our
proposed network over other state-of-the-art steganalyzers either
hand-crafted or learned using deep networks in the literature.


The rest of the paper is organized as follows. In
Sect.~\ref{sec:framework_all}, we firstly show the theoretical
rationale of our proposed WISERNet, and then describe the detailed
structure of WISERNet. Results of experiments conducted on different
color image datasets are presented in Sect.~\ref{sec:exp}. Finally, we
make conclusions in Sect.~\ref{sec:conclude}.

\section{Our proposed WISERNet}
\label{sec:framework_all}

In this section, we firstly introduce convolutional layers, the
principal part of CNN as preliminaries. Then we discuss the motivation
of our proposed WISERNet in theory. Finally we provide the conceptual
architecture of WISERNet, as well as its detailed configuration.
\subsection{Preliminaries}
\label{sec:preliminaries}
In this paper we only consider RGB true-color model. Given
$\textrm{\textbf{X}}$, a true-color image of size of $M \times N$, it
comprises three bands, namely the \textit{red}, the \textit{green},
and the \textit{blue} band. In our research, we do not take the
specific characteristic of a band into consideration. Therefore
without loss of generality, $\textrm{\textbf{X}}$ can be represented
as
$\{\textrm{\textbf{X}}_{1},\textrm{\textbf{X}}_{2},\textrm{\textbf{X}}_{3}\}$,
where $\textrm{\textbf{X}}_{i}=(x_{i,pq})^{M \times N}$,
$x_{i,pq} \in \{ 0,1,\cdots,255\}$, $1 \le i \le 3$, $1 \le p \le M$
and $1 \le q \le N$. All of the state-of-the-art steganographic
algorithms which can be applied on color
images~(\hspace{1sp}\cite{holub_eurasip_2014,li_icip_2014,
  denemark_ihmmsec_2015,li_tifs_2015,tang_spl_2016}) add zero-mean
$\pm 1$ additional stego noise to every target band of a given
cover image. Therefore, every band in $\textrm{\textbf{X}}$ can be
further represented as
$\textrm{\textbf{X}}_{i}=(c_{i,pq}+n_{i,pq})^{M \times
  N}=\textrm{\textbf{C}}_{i}+\textrm{\textbf{N}}_{i}$,
where $\textrm{\textbf{C}}_{i}=(c_{i,pq})^{M \times N}$,
$c_{i,pq} \in \{ 0,1,\cdots,255\}$ denotes the corresponding cover
band, and
$\textrm{\textbf{N}}_{i}=(n_{i,pq})^{M \times N}, n_{i,pq} \in \{
-1,0,+1\}$
denotes the additive stego noise matrix added to
$\textrm{\textbf{X}}_{i}$. For an innocent cover image,
$\textrm{\textbf{N}}_{i}$, $1 \le i \le 3$ are all zero matrices.

All of the state-of-the-art deep-learning
steganalyzers~\cite{tan_apsipa_2014,qian_spie_2015,xu_spl_2016,
  pibre_ei_2016_ver2,qian_icip_2016, zeng_tifs_2018,
  zeng_ei_2017,xu_ihmmsec_2017,chen_ihmmsec_2017,ye_tifs_2017} are
based on CNN~(Convolutional Neural Network). The principal part of CNN
is a cascade of alternating convolutional layers, regulation
layers~(e.g. BN layers~\cite{ioffe_icml_2015}) and pooling layers. On
top of the principal part, there are optional multiple fully-connected
layers. Convolutional layers are the core building blocks of a
CNN. For a given convolutional layer $L_l$, it takes $J$ channels of
inputs $\textrm{\textbf{Z}}^{l-1}_{j}$, $1 \le j \le J$, convolves
them with an array of $J \times K$ kernels
$\textrm{\textbf{W}}^{l}_{jk}$~(usually with learnable weights) and
generates $K$ channels of outputs $\textrm{\textbf{Z}}^{l}_{k}$,
$1 \le k \le K$. In the context of image steganalysis,
$\textrm{\textbf{Z}}^{l-1}_{j}$, $\textrm{\textbf{Z}}^{l}_{k}$, and
$\textrm{\textbf{W}}^{l}_{jk}$ are always represented as
two-dimensional matrices. The \textit{normal convolution} can be
modeled as:
\begin{equation}
  \label{eq:normal_conv}
  \textrm{\textbf{Z}}^{l}_{k}=\sum^{J}_{j=1}\textrm{\textbf{Z}}^{l-1}_{j} \ast \textrm{\textbf{W}}^{l}_{jk}, 1 \le k \le K
\end{equation}

We utilize \textit{channel-wise convolution}, a variant of normal
convolution in our work.  In channel-wise convolution, each input
channel corresponds to standalone $K$ output channels and is convolved
with an array of $K$ kernels. As a result with $J$ input channels we
can get $J \times K$ output channels:
\begin{multline}
  \label{eq:channel_wise_conv}
  \qquad\qquad\textrm{\textbf{Z}}^{l}_{k'}=\textrm{\textbf{Z}}^{l-1}_{j} \ast \textrm{\textbf{W}}^{l}_{jk}, \\
  k'=(j-1)\times K+k, 1 \le j \le J, 1 \le k \le K
\end{multline}

The existing deep-learning
steganalyzers~\cite{tan_apsipa_2014,qian_spie_2015,xu_spl_2016,
  pibre_ei_2016_ver2,qian_icip_2016,
  zeng_tifs_2018,zeng_ei_2017,xu_ihmmsec_2017,chen_ihmmsec_2017,ye_tifs_2017}
incorporate the domain knowledge behind rich models, and initialize
the kernels in the bottom convolutional layer as high-pass filters to
increase SNR~(Signal-to-Noise Ratio)~\footnote{In the literature of
  steganalysis, image content is ``noise'' while stego noise is
  ``signal''.}. Fixed weights in the bottom kernels are adopted in
most existing deep-learning
steganalyzers~\cite{tan_apsipa_2014,qian_spie_2015,pibre_ei_2016_ver2,qian_icip_2016,xu_spl_2016,chen_ihmmsec_2017,xu_ihmmsec_2017,zeng_ei_2017,zeng_tifs_2018},
while learnable weights are adopted in Ye's
model~\cite{ye_tifs_2017}.

\subsection{Rationale of our proposed WISERNet}
\label{sec:rationale}

\begin{table}[!t]
  \centering
  \caption[]{ Means of the absolute value of the correlation
    between the intensity values of different color
    bands versus those of the
    corresponding stego noises. The results on
    BOSS-PPG-LAN for HILL, CMD-C-HILL, SUNIWARD, and
    CMD-C-SUNIWARD with 0.4 bpc payload are reported.}
  \resizebox{0.95\columnwidth}{!}{%
    {\renewcommand{\arraystretch}{1.2}
      \begin{tabular}{cccc}
        \Xhline{2\arrayrulewidth}
        \textbf{Elements} & \textbf{\makecell{\textit{red} band \\
        vs. \\\textit{green} band}} & \textbf{\makecell{\textit{red} band \\
        vs. \\\textit{blue} band}} & \textbf{\makecell{\textit{blue} band \\
        vs. \\\textit{green} band}} \\
        \hline
        \multicolumn{4}{ c }{\textbf{For HILL steganography}} \\
        \hline
        Intensity values & 0.9317    & 0.8297    & 0.9217 \\
        Stego noises & 0.0024    & 0.0023    & 0.0024 \\
        \hline
        \multicolumn{4}{ c }{\textbf{For CMD-C-HILL steganography}} \\
        \hline
        Intensity values & 0.9317    & 0.8297    & 0.9217 \\
        Stego noises & 0.2704    & 0.2619    & 0.2859 \\
        \hline
        \multicolumn{4}{ c }{\textbf{For SUNIWARD steganography}} \\
        \hline
        Intensity values & 0.9317    & 0.8297    & 0.9217 \\
        Stego noises & 0.0022    & 0.0021    & 0.0022 \\
        \hline
        \multicolumn{4}{ c }{\textbf{For CMD-C-SUNIWARD steganography}} \\
        \hline
        Intensity values & 0.9317    & 0.8297    & 0.9217 \\
        Stego noises & 0.2980    & 0.2817    & 0.2975 \\
        \Xhline{2\arrayrulewidth}
      \end{tabular}
    }}
  \label{tab:avg_abs_corrs_intensities_stego_noises}
\end{table}

\begin{table*}[!t]
  \centering
  \caption[]{Impact of growing
    $\rho_{\textrm{r-g}}=\rho_{\textrm{r-b}}=\rho$ on
    $\frac{\textrm{MMD}_c}{\textrm{MMD}_n}$.}
  \resizebox{0.75\textwidth}{!}{%
    {\renewcommand{\arraystretch}{1.5}
      \begin{tabular}{ccccccccccc}
        \Xhline{2\arrayrulewidth}
        \multicolumn{11}{ c }{\textbf{$\rho_{\textrm{r-g}}=\rho_{\textrm{r-b}}=\rho$}}\\
        \hline
        0 & 0.1 & 0.2 & 0.3 & 0.4 & 0.5 & 0.6 & 0.7 & 0.8 & 0.9 & 1 \\
        \hline
        1.178 & 1.158  & 1.124  & 1.108  & 1.064  & 1.038  & 1.027  & 1.021  & 1.013  & 1.008  & 1.000  \\
        \Xhline{2\arrayrulewidth}
      \end{tabular}
    }
  }
  \label{tab:mmd_increasing_corr}
\end{table*}

If we apply normal convolution with $K$ output channels to a
true-color image in the bottom convolutional layer, we get:
\begin{IEEEeqnarray}{rl}
  \label{eq:normal_conv_collusion}
  & \textrm{\textbf{Z}}^{1}_{k}=\sum^{3}_{j=1}\textrm{\textbf{X}}_{j} \ast \textrm{\textbf{W}}^{1}_{jk},\ 1 \le k \le K     \IEEEnonumber\\
  = & \sum^{3}_{j=1}\textrm{\textbf{C}}_{j} \ast
  \textrm{\textbf{W}}^{1}_{jk}+\sum^{3}_{j=1}\textrm{\textbf{N}}_{j}
  \ast \textrm{\textbf{W}}^{1}_{jk},\ 1 \le k \le K
\end{IEEEeqnarray} 

Our rationale starts from the following statement about existing
steganographic algorithms for color images:

For a true-color stego image, the intensity values in the same
location of the three bands exhibit strong correlation. Their
means~(or expectations) are similar from the perspective of
statistics. Conversely, for steganographic algorithms originally
oriented to gray-scale images,
e.g. \cite{holub_eurasip_2014,li_icip_2014,
  sedighi_tifs_2016,denemark_ihmmsec_2015,li_tifs_2015}, the zero-mean
$\pm 1$ additional stego noises in the same location of the three
bands exhibit no correlation. Even for \cite{tang_spl_2016} which is
committed to increase the correlation of the stego noises among bands,
they still exhibit weak correlation. This is because the
well-established rate-distortion bound~\cite{filler_tifs_2010}
determines that strong correlation among stego noises and minimal
possible distortion are in conflict.

To verify the above statement, we analyzed the $10,000$
BOSS-PPG-LAN~(see Sect.~\ref{sec:setups}) stego images generated by
HILL, CMD-C-HILL, SUNIWARD, and CMD-C-SUNIWARD, respectively, all with
0.4 bpc~(bits per channel/band pixel) embedding rate. In the
experiment, means of the absolute value of the correlation between the
intensity values of different color bands were calculated. We compared
them with those of the corresponding stego noises. From
Tab.~\ref{tab:avg_abs_corrs_intensities_stego_noises}, we can see that
for all of the four steganographic algorithms, stego noises have no
effect on the correlation of the intensity values among bands. They
all exhibited notably strong correlation. On the other hand, for HILL
and SUNIWARD, the stego noises among bands exhibit nearly zero
correlation. Even for CMD-C-HILL and CMD-C-SUNIWARD, they exhibit weak
correlation.

As mentioned in Sect.~\ref{sec:preliminaries}, the main purpose of the
convolutions in the bottom convolutional layer is to boost SNR, namely
suppress image contents~(intensity values) and retain stego noises at
the same time. Let $\textrm{E}([\bullet])$ and
$\textrm{Var}([\bullet])$ denote the expectation and the variance of
the elements in matrix $[\bullet]$;
$\textrm{Corr}([\bullet],[\square])$ denotes the correlation of the
corresponding elements in matrix $[\bullet]$ and
$[\square]$. Following the convention in image
processing~\cite{gonzalez_dip_2002}, we define SNR of a given
two-dimensional input $[\bullet]$ as:
\begin{equation}
  \label{eq:snr_define}
  \textrm{SNR}([\bullet])=\frac{\textrm{Var}([\bullet])}{\textrm{E}^2([\bullet])}
\end{equation}
According to the linearity of expectation and
\eqref{eq:normal_conv_collusion}, for a given
$\textrm{\textbf{Z}}^{1}_{k}$ we can get:
\begin{equation}
  \label{eq:snr_Z_1k}
  \textrm{SNR}(\textrm{\textbf{Z}}^{1}_{k})=
  \frac{\textrm{Var}(\sum^{3}_{j=1}\textrm{\textbf{N}}_{j} \ast
    \textrm{\textbf{W}}^{1}_{jk})}{\textrm{E}^2(\sum^{3}_{j=1}\textrm{\textbf{C}}_{j}
    \ast
    \textrm{\textbf{W}}^{1}_{jk})} = \frac{\textrm{Var}(\sum^{3}_{j=1}\textrm{\textbf{N}}_{j} \ast
    \textrm{\textbf{W}}^{1}_{jk})}{(\sum^{3}_{j=1}\textrm{E}(\textrm{\textbf{C}}_{j}
    \ast \textrm{\textbf{W}}^{1}_{jk}))^2}
\end{equation}

Initially, we set
$\textrm{\textbf{W}}^{1}_{1k}=\textrm{\textbf{W}}^{1}_{2k}=\textrm{\textbf{W}}^{1}_{3k}=\widetilde{\textrm{\textbf{W}}}_k$,
where $\widetilde{\textrm{\textbf{W}}}_k$ is a predefined high-pass
filter.  Since
$\textrm{E}(\textrm{\textbf{C}}_{1}) \approx
\textrm{E}(\textrm{\textbf{C}}_{2}) \approx
\textrm{E}(\textrm{\textbf{C}}_{3})$,
we can get
$\textrm{E}(\textrm{\textbf{C}}_{1} \ast \textrm{\textbf{W}}^{1}_{1k})
\approx \textrm{E}(\textrm{\textbf{C}}_{2} \ast
\textrm{\textbf{W}}^{1}_{2k}) \approx
\textrm{E}(\textrm{\textbf{C}}_{3} \ast
\textrm{\textbf{W}}^{1}_{3k})=\mu$~(as
demonstrated later in Tab.~\ref{tab:avg_corr_vs_iterations} of
Sect.~\ref{sec:impact_learnable_bottom_kernels}, the assumption still holds along with
increasing training iterations of WISERNet). Denote
$\textrm{Var}(\textrm{\textbf{N}}_{j} \ast
\textrm{\textbf{W}}^{1}_{jk})=\sigma_{j}^{2},\ 1 \le j \le 3$.
Existing color image steganography tends to uniformly distribute
embedding changes to three color bands, therefore for the sake of
simplicity, we further assume that
$\sigma_{j}=\sigma,\ 1 \le j \le 3$.  From \eqref{eq:snr_Z_1k} we can
get:
\begin{equation}
  \label{eq:snr_Z_1k_deduce}
  \textrm{SNR}(\textrm{\textbf{Z}}^{1}_{k})=\frac{\sum^{3}_{j=1}\textrm{Var}(\textrm{\textbf{N}}_{j} \ast
    \textrm{\textbf{W}}^{1}_{jk})+\Delta}{9\mu^2}=\frac{\sum^{3}_{j=1}\sigma_{j}^{2}+\Delta}{9\mu^2}
\end{equation}
in which:
\begin{equation}
  \label{eq:delta_cov_to_corr}
  \Delta= 2\cdot\sum_{1 \le i < j \le 3}\textrm{Corr}(\textrm{\textbf{N}}_{i}
  \ast \textrm{\textbf{W}}^{1}_{ik},\textrm{\textbf{N}}_{j} \ast
  \textrm{\textbf{W}}^{1}_{jk})\cdot\sigma_{i}\sigma_{j}
\end{equation} 
Please note that a discrete convolution is a linear transform, and a
linear transform will never change the correlation between random
variables~\cite{carey_conv_corr}. Therefore:
\begin{equation}
  \label{eq:delta_corr}
  \textrm{Corr}(\textrm{\textbf{N}}_{i}
  \ast \textrm{\textbf{W}}^{1}_{ik},\textrm{\textbf{N}}_{j} \ast
  \textrm{\textbf{W}}^{1}_{jk})=\textrm{Corr}(\textrm{\textbf{N}}_{i},\textrm{\textbf{N}}_{j}),\ 1 \le i < j \le 3
\end{equation}

From \eqref{eq:delta_cov_to_corr} and \eqref{eq:delta_corr} we can set
$\Delta=0$ for HILL and SUNIWARD since the inter-band correlation
of the stego noises generated by them is nearly zero. As a result:
\begin{equation}
  \label{eq:eq:snr_Z_1k_final_hill_suniward}
    \textrm{SNR}(\textrm{\textbf{Z}}^{1}_{k})=\frac{\sum^{3}_{j=1}\sigma_{j}^{2}}{9\mu^2}=\frac{1}{3}
    \cdot \frac{\sigma^2}{\mu^2}
\end{equation}

For CMD-C, $\Delta$ is unneglectable. However, the inter-band
correlation of the stego noises stays weak. Using the statistics
reported in Tab.~\ref{tab:avg_abs_corrs_intensities_stego_noises}, we
set
$\textrm{Corr}(\textrm{\textbf{N}}_{i},\textrm{\textbf{N}}_{j})\approx
0.3,\ 1 \le i < j \le 3$. From
\eqref{eq:snr_Z_1k_deduce} we can get:
\begin{equation}
  \label{eq:eq:snr_Z_1k_final_cmdc}
  \textrm{SNR}(\textrm{\textbf{Z}}^{1}_{k})\approx\frac{3\sigma^2+2\cdot3\cdot0.3\cdot\sigma^2}{9\mu^2}=\frac{5}{9}
  \cdot \frac{\sigma^2}{\mu^2}
\end{equation}

Please note that:
\begin{equation} 
  \label{eq:snr_X_j_conv_W1jk}
  \textrm{SNR}(\textrm{\textbf{X}}_{j} \ast \textrm{\textbf{W}}^{1}_{jk})=\frac{\textrm{Var}(\textrm{\textbf{N}}_{j} \ast \textrm{\textbf{W}}^{1}_{jk})}{\textrm{E}^2(\textrm{\textbf{C}}_{j} \ast
    \textrm{\textbf{W}}^{1}_{jk})}=\frac{\sigma^2}{\mu^2},\ 1 \le j \le 3
\end{equation}
Compare \eqref{eq:eq:snr_Z_1k_final_hill_suniward},
\eqref{eq:eq:snr_Z_1k_final_cmdc} with \eqref{eq:snr_X_j_conv_W1jk},
we can see for those steganographic algorithms applied on color
images,
$\textrm{SNR}(\textrm{\textbf{Z}}^{1}_{k})<\textrm{SNR}(\textrm{\textbf{X}}_{j}
\ast \textrm{\textbf{W}}^{1}_{jk}),\ 1 \le j \le 3$.
The summation in $\textrm{\textbf{Z}}^{1}_{k}$ actually impairs the
SNR which has been boosted by the convolutions in
$\textrm{\textbf{X}}_{j} \ast \textrm{\textbf{W}}^{1}_{jk},\ 1 \le j
\le 3$.
In fact, we can regard the summation in normal convolution one sort of
``linear collusion attack''~\cite{su_tm_2005} which is the process of
forming a linear combination of input bands, and as a result can
reserves strong correlated patterns, while impairs uncorrelated
noises~(or weak correlated signals) in input bands. Accordingly, we
decide not to apply normal convolution in the bottom convolutional
layer.

\begin{figure*}[!t]
  \centering
  \includegraphics[width=0.75\linewidth,keepaspectratio]{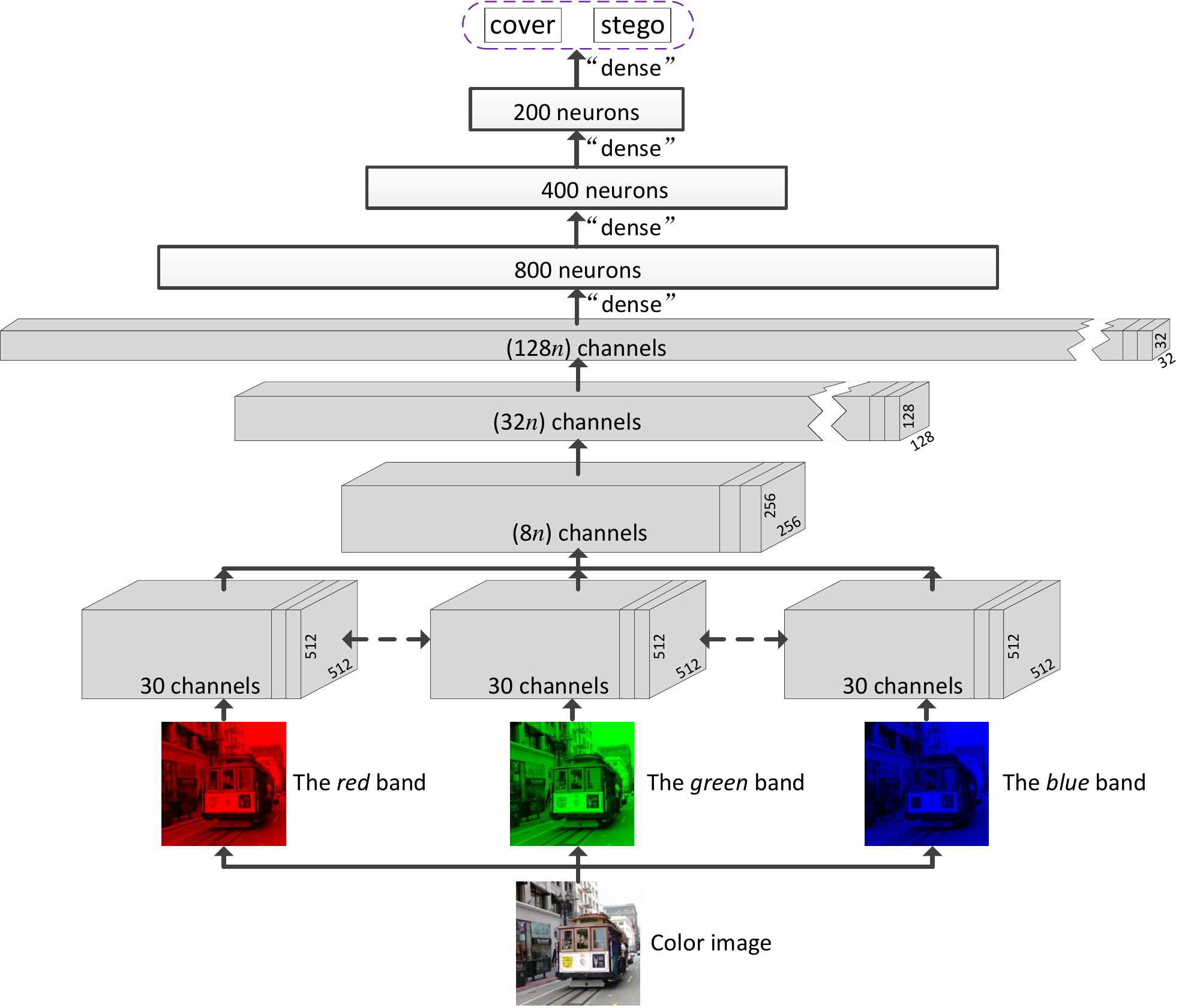} 
  \caption[]{Conceptual architecture of our proposed WISERNet.}
  \label{fig:conceptual_arch}  
\end{figure*} 

In practice, there are three solutions to bypass the summation in normal
convolution. The first solution is to directly concatenate the three
bands of $\textrm{\textbf{X}}$ to generate a one-band input
$\textrm{\textbf{X}}'$ for the bottom convolutional layer, which is
straightforward:
\begin{multline}
  \label{eq:def_x_p}
  \qquad\textrm{\textbf{X}}'=(x^{'}_{pq'})^{M \times 3N},\\
  x^{'}_{pq'}=x_{i,pq},\ 
  q'=(i-1)\times N+q,\\
  1 \le i \le 3,\ 1 \le p \le M,\ 1 \le q \le N
\end{multline}
The second solution is to interleave the intensity values from three
bands, e.g. in a fashion that
$(x_{1,pq},\ x_{2,pq},\ x_{3,pq},\ x_{1,(p+1)q},\ x_{2,(p+1)q},\
x_{3,(p+1)q},\ \cdots)$.
The third solution is channel-wise convolution. We adopt channel-wise
convolution in our proposed network since we believe the first two
solutions are inferior to the third solution. The argument is as
follows.

Given $(x_{1,pq},x_{2,pq},x_{3,pq})$, three intensity values in the
same location of the three color bands. We have already known that
they exhibit strong correlation. If we adopt the first solution,
a.k.a. the straightforward solution, then
$(x_{1,pq},x_{2,pq},x_{3,pq})$ corresponds to
$(x^{'}_{pq},x^{'}_{p(N+q)},x^{'}_{p(2N+q)})$ in
$\textrm{\textbf{X}}'$. Please note that the column distance of
$(x^{'}_{pq},x^{'}_{p(N+q)},x^{'}_{p(2N+q)})$ is $N$, far beyond the
usual perception field of the kernels in lower convolutional
layers. As a result, the originally strong correlation in
$(x_{1,pq},x_{2,pq},x_{3,pq})$, after mapped to
$(x^{'}_{pq},x^{'}_{p(N+q)},x^{'}_{p(2N+q)})$, cannot be catched in
lower convolutional layers. Only convolution kernels in the top layers
of very deep networks can perceive it. The second solution is also
infeasible. You can think of it as a noisy up-sampling procedure of a
given band in which the step size is two. Those weak $\pm 1$ stego
noises will be by large concealed under the relatively more powerful
up-sampling noises.

On the other hand, if we adopt channel-wise convolution, then after
convolution, those elements affected by $(x_{1,pq},x_{2,pq},x_{3,pq})$
are still allocated at the corresponding locations of the output
channels. The strong correlation in them is thus kept, and can be
perceived by all of the convolutional layers from bottom to top.

In order to verify that the bottom channel-wise convolutional layer do
help to preserve stego noises and consequently boost the SNR, we
conducted an evaluation experiment as follows:

Firstly, 40\% pixels of every band of every cover image in a dataset
used in our experiments~(BOSS-PPG-LAN, please refer to
Sect.~\ref{sec:setups}) are stochastically selected to perform random
$\pm 1$ modifications. The modifications simulate the effect of
na\"{\i}ve LSB matching embedding. Then we apply normal convolution or
channel-wise convolution to the cover images and their corresponding
pseudo-stego images. Only $\mathbf{K}_5$, one of the 30 spatial-domain
rich model kernels~\cite{fridrich_tifs_2012}, is used in the
convolution to further reduce the complexity. With $\mathbf{K}_5$,
whether normal convolution or channel-wise convolution generates a
single output feature map. Next, we extract the 686-dimensional
SPAM~(Subtractive Pixel Adjacency Matrix)~\cite{pevny_tifs_2010}
steganalytic feature vector for every output feature map.  Finally, we
compute the MMD~(Maximum Mean Discrepancy)~\cite{pevny_ih_2008}
between the cover images and the corresponding pseudo-stego images in
the SPAM feature space. The presence of a higher MMD value indicates
that it is easier to distinguish stego images from cover images.

Please note that a given stego image can be considered to be a noisy
version of the corresponding cover image, and $\mathbf{K}_5$ is a
high-pass filter aims at boosting the SNR. Therefore with the same
$\mathbf{K}_5$ kernel, the convolution type with relatively higher MMD
value helps to preserve stego noises. Let $\textrm{MMD}_n$ and
$\textrm{MMD}_c$ denotes the MMD value for the output feature maps
generated by normal convolution and channel-wise convolution,
respectively. We can get $\textrm{MMD}_n=0.01187$ while
$\textrm{MMD}_c=0.01398$. As a result
$\frac{\textrm{MMD}_c}{\textrm{MMD}_n}=1.178$, which implies that the
bottom channel-wise convolutional layer do help to preserve stego
noises and consequently boost SNR when stego noises in separate color
bands exhibit no correlation.

Furthermore, we investigate the impact of growing correlation of stego
noises in different bands on
$\frac{\textrm{MMD}_c}{\textrm{MMD}_n}$. For every cover image, we
firstly stochastically select 40\% pixels of the \textit{red} band to
perform random $\pm 1$ modifications. Let $\rho_{\textrm{r-g}}$
denotes the correlation of the $\pm 1$ modifications in the
\textit{red} band and those in the \textit{green} band. Analogically,
The meaning of $\rho_{\textrm{r-b}}$ and $\rho_{\textrm{g-b}}$ are
self-evident.  The $\pm 1$ modifications are performed on the
\textit{green} band and the \textit{blue} band, to guarantee that
$\rho_{\textrm{r-g}}=\rho_{\textrm{r-b}}=\rho,\ 0 < \rho \le 1$. It is
easy to verify that $2\rho^2-1 \le \rho_{\textrm{g-b}} \le 1$ with
$\rho_{\textrm{r-g}}=\rho_{\textrm{r-b}}=\rho$, and
$\rho_{\textrm{g-b}} \to 1$ along with $\rho \to 1$. Therefore as a
compromise, we only investigate the impact of growing
$\rho_{\textrm{r-g}}=\rho_{\textrm{r-b}}=\rho$ on
$\frac{\textrm{MMD}_c}{\textrm{MMD}_n}$. As shown in
Tab.~\ref{tab:mmd_increasing_corr},
$\frac{\textrm{MMD}_c}{\textrm{MMD}_n}$ keeps reducing along with
increasing $\rho$, which implies that the gain introduced by
channel-wise convolution gradually decreases with increasing
inter-band correlation of stego noises. However, since even for the
state-of-the-art steganographic algorithms~(e.g. CMD-C) taking into
account the correlation of stego noises among bands, the inter-band
correlation of stego noises still keeps weak~(is approximately equal
to 0.3, as shown in
Tab.~\ref{tab:avg_abs_corrs_intensities_stego_noises}). Therefore even
for those algorithms like CMD-C, the gain introduced by channel-wise
convolution still cannot be neglected.

We decide to only adopt channel-wise convolution and withdraw
summation in the bottom convolutional layer. In the upper
convolutional layers we still adopt normal convolution which retains
summation. This is because withdrawal of the summation in convolution
is a double-edged sword. Please note that during the generation
procedure of true-color images, various kinds of sources, especially
the CFA interpolation algorithm, introduces dependencies among pixels
and bands. The dependencies can be regarded as hidden patterns stay
approximately the same in color bands. Though those hidden patterns
are severely weaken in varying post-processing procedures, it is still
potential to be catched by deep-learning networks. In a CNN based
deep-learning network, since we have already known discrete
convolution will never change the correlation between input channels,
for a given convolutional layer $L_l$ we can also get:
\begin{equation}
  \label{eq:z_corr}
  \textrm{Corr}(\textrm{\textbf{Z}}^{l-1}_{i}
  \ast \textrm{\textbf{W}}^{l}_{ik},\textrm{\textbf{Z}}^{l-1}_{j} \ast
  \textrm{\textbf{W}}^{l}_{jk})=\textrm{Corr}(\textrm{\textbf{Z}}^{l-1}_{i},\textrm{\textbf{Z}}^{l-1}_{j})
\end{equation}
Therefore if we model the hidden patterns a certain sort of
correlation among bands, it can be passed forward through the
convolutional layers from bottom to top, even though disturbed by the
nonlinearities in the pipeline. Since every normal convolutional layer
can regarded as one sort of ``linear collusion attack'', the
summations in cascaded convolutional layers may enhance the ability of
``hidden-pattern-aware'' of our proposed network. Therefore only in
the bottom convolutional layer, which aims at suppressing contents
while retaining stego noises, we withdraw summation and adopt
channel-wise convolution.

It has been verified that in ``linear collusion attack'', the more the
objects involved in collusion, the better the strong correlated
patterns are reserved, the worse the weak correlated signals are
impaired~\cite{su_tm_2005}. Since we regard the summation in normal
convolution as one sort of ``linear collusion attack'', it is clear
that the more the kernels in a given normal convolutional layer, the
more the outputs can be involved in summation, and therefore the
better results of the ``linear collusion attack'' we can expect. Based
on the above analysis, our design concept is different from typically
making the deep-learning network deeper. We try to promote the
performance of our proposed steganalyzer by making the upper
convolutional layers wider, namely expanding the number of their
convolution kernels.

\subsection{WISERNet: the wider separate-then-reunion network}
\label{sec:framework}

In light of the rationale presented in Sect.~\ref{sec:rationale}, we
propose WISERNet, the WIder SEparate-then-Reunion Network.  The
conceptual architecture of WISERNet is illustrated in
Fig.~\ref{fig:conceptual_arch}.  

WISERNet takes a true-color image as input and applies channel-wise
convolution to the red, green, and blue bands of the input image,
respectively. Following the recipe of prior
works~\cite{tan_apsipa_2014,qian_spie_2015,xu_spl_2016,
  pibre_ei_2016_ver2,qian_icip_2016,
  zeng_tifs_2018,zeng_ei_2017,xu_ihmmsec_2017,chen_ihmmsec_2017,ye_tifs_2017},
the weights of the kernels assigned to each channel are initialized
with high-pass filters in rich models to increase SNR. Here the thirty
filters used in SRM~\cite{fridrich_tifs_2012} are used. Therefore each
band is convolved with thirty $5 \times 5$ initialized kernels and
thirty corresponding output channels are generated.  Please note that
in the training procedure, we set the weights in the bottom kernels
\textit{learnable}. The bottom channel-wise convolutional layer
corresponds to the ``separate'' stage of our proposed network. The
three separate groups of output channels are then concatenated
together to form a ninety-channel input of the second convolutional
layer.

Started from the second convolutional layer, the upper
structure of our proposed network corresponds to the ``reunion''
stage. It is a united wide and relatively shallow convolutional neural
network, which contains convolutional layers with plenty of
kernels. We fix the depth, namely the number of cascaded convolutional
layers of the upper structure to three, and explicitly increase the
capacity of WISERNet by magnifying the number of the kernels in each
convolutional layer with a model magnification factor $\bm{n}$. With
increasing $\bm{n}$, WISERNet becomes ``wider'' and ``wider''. On top
of the convolutional layers there is a four-layer
fully-connected~(``dense'') neural network which makes the final
prediction. The successive layers of the fully-connected network
contain 800, 400, 200, and 2 neurons, respectively.

\begin{table}[!t]
  \centering
  \caption[]{The detailed configuration of our proposed WISERNet.}
  \resizebox{\columnwidth}{!}{%
    {\renewcommand{\arraystretch}{1.2}
      \begin{tabular}{cccc}
        \Xhline{2\arrayrulewidth}
        \textbf{Type} & \makecell{\textbf{Kernels size/stride} \\ {\footnotesize \textit{(width$\times$height$\times$depth)$/$stride}}} & \makecell{\textbf{Output size} \\ {\footnotesize \textit{(width$\times$height$\times$channel)}}} \\
        \hline
        \makecell{Channel-wise \\ convolution} & $(5\times 5 \times 30)/1$ & $512 \times 512 \times 90$\\ 
        \hline
        Convolution & $(5\times 5 \times  \bm{(8n)})/2$ & $256 \times 256 \times \bm{(8n)}$\\
        \setarstrut{\tiny}%
        {\tiny ABS} & {\tiny /} & {\tiny ------}\\
        {\tiny BN} & {\tiny /} & {\tiny ------}\\
        {\tiny ReLU} & {\tiny /} & {\tiny ------}\\
        \restorearstrut
        Average pooling & $(5\times 5 \times 1)/2$ & $128 \times 128 \times$ $\bm{(8n)}$\\
        \hline
        Convolution & $(3\times 3 \times \bm{(32n)})/1$ & $128 \times 128 \times \bm{(32n)}$\\
        \setarstrut{\tiny}%
        {\tiny BN} & {\tiny /} & {\tiny ------}\\
        {\tiny ReLU} & {\tiny /} & {\tiny ------}\\
        \restorearstrut
        Average pooling & $(5\times 5 \times 1)/4$ & $32 \times 32 \times \bm{(32n)}$\\
        \hline
        Convolution & $(3\times 3 \times \bm{(128n)})/1$ & $32 \times 32 \times \bm{(128n)}$\\
        \setarstrut{\tiny}%
        {\tiny BN} & {\tiny /} & {\tiny ------}\\
        {\tiny ReLU} & {\tiny /} & {\tiny ------}\\
        \restorearstrut
        Average pooling & $(32\times 32 \times 1)/32$ & $1 \times 1 \times \bm{(128n)}$\\
        Flatten & / & $\bm{128n}$\\
        \hline
        Fully connection & / & $\bm{800}$\\
        \setarstrut{\tiny}%
        {\tiny ReLU} & {\tiny /} & {\tiny ------}\\
        \restorearstrut
        Fully connection & / & $\bm{400}$\\
        \setarstrut{\tiny}%
        {\tiny ReLU} & {\tiny /} & {\tiny ------}\\
        \restorearstrut
        Fully connection & / & $\bm{200}$\\
        \setarstrut{\tiny}%
        {\tiny ReLU} & {\tiny /} & {\tiny ------}\\
        \restorearstrut
        Fully connection & / & $\bm{2}$\\
        \setarstrut{\tiny}%
        {\tiny Softmax} & {\tiny /} & {\tiny ------}\\
        \restorearstrut
        \Xhline{2\arrayrulewidth}
      \end{tabular}
    }}
  \label{tab:detailed_configuration}
\end{table}

The detailed configuration of WISERNet is shown in
Tab.~\ref{tab:detailed_configuration}. Assume that the true-color
image fed to WISERNet is of size $512 \times 512$.  From
Tab.~\ref{tab:detailed_configuration} we can see the output of the
bottom channel-wise convolutional layer, the ``separate'' stage, is
ninety channels of feature maps of size $512 \times 512$, which act as
the input of the ``reunion'' stage. In the ``reunion'' stage, all of
the normal convolutional layers are followed by a BN~(Batch
Normalization) layer, a ReLU~(Rectified Linear Unit) layer, and an
average pooling layer successively. Specifically, the absolute values
of the output of the first normal convolutional layer are fed forward,
following the recipe of prior
works~\cite{ye_tifs_2017,zeng_tifs_2018,zeng_ei_2017,xu_ihmmsec_2017,chen_ihmmsec_2017}.
The size of the output feature maps of the normal convolutional layers
from bottom to top in this stage is $256 \times 256$,
$128 \times 128$, and $32 \times 32$, respectively. In order to
roughly preserve the time complexity per layer, the number of the
kernels in each convolutional layer is quadrupled
accordingly. Consequently, the number of the output feature maps is
$\bm{8n}$, $\bm{32n}$, and $\bm{128n}$ with the model magnification
factor $\bm{n}$, respectively. Ahead of the top-most normal
convolutional layer, the output feature maps are pooled with a large
stride~(step=32) and then flatten to a $\bm{128n}$-D feature vector,
which further acts as the input of the top fully-connected network. In
the top fully-connected network, ReLU activation functions are used in
all three hidden layers. The final layer contains two neurons which
denote ``stego'' prediction and ``cover'' prediction. Softmax function
is used to output predicted probabilities.

\section{Experiments}
\label{sec:exp}

\subsection{Experiment setup}
\label{sec:setups}

As reported in prior
works~\cite{goljan_wifs_2014,goljan_spie_2015,abdulrahman_iwcc_2015,abdulrahman_scn_2016,abdulrahman_ihmmsec_2016},
different CFA demosaicking algorithms and different down-sampling
algorithms greatly affect detection performance of existing color
image steganalyzers. Therefore, all experiments in this paper are
conducted on different versions of
BOSSBase~(v1.01)~\cite{bas_ih_2011}. Starting with the 10,000
full-resolution raw images, firstly we followed the dataset generating
process used in \cite{goljan_wifs_2014}. We used ufraw to demosaick
the raw images and then used ImageMagick ``convert'' utility with the
default ``Lanczos'' kernel to down-sample~(set the smaller image
dimension to 512) and central crop the resulting PPM color images to
$512 \times 512$. We used two demosaicking algorithms in ufraw,
PPG~(Patterned Pixel Grouping) and AHD~(Adaptive Homogeneity
Directed). The corresponding datasets were named BOSS-PPG-LAN and
BOSS-AHD-LAN. Following the same process as above, with the
down-sampling operation removed, we obtained another two datasets,
BOSS-PPG-CRP and BOSS-AHD-CRP. In order to explore the impact of
different down-sampling operations on detection performance of color
image steganalyzers, we replaced the ``convert'' utility with
correspondent Matlab$^{\textrm{\textregistered}}$ script. Pairing PPG,
or AHD demosaicking algorithm with ``Bicubic'', or ``Bilinear'' kernel
in Matlab$^{\textrm{\textregistered}}$ function \textit{imresize}, we
generated four more datasets, BOSS-PPG-BIC, BOSS-PPG-BIL, BOSS-AHD-BIC
and BOSS-AHD-BIL, respectively.

Four color image steganographic algorithms, HILL, SUNIWARD,
CMD-C-HILL, and CMD-C-SUNIWARD were our attacking targets in the
experiments.~\footnote{According to peer feedback, we have fixed a bug
  in the original implementation of CMD-C~\cite{tang_spl_2016} and
  guarantee that different pseudo-random seeds are assigned to each of
  the simulators corresponding to three color bands.} Their embedding
payloads were set to 0.1, 0.2, 0.3, 0.4, and 0.5 bpc, in which we
mainly focused on 0.2 bpc and 0.4 bpc. For HILL and SUNIWARD, the same
payload was embedded in every band.

\begin{table}[!t]
  \centering
  \caption[]{Detection performance of CRM, SGRM, and GCRM on four
    different datasets. In each sub-table, the best results for 0.2 bpc
    are underlined, while the best results for 0.4 bpc are in framed
    boxes. }
  \resizebox{\columnwidth}{!}{%
    {\renewcommand{\arraystretch}{1.4}
      \begin{tabular}{ccccccc}
        \Xhline{2\arrayrulewidth}
        \multirow{3}{*}{\large\textbf{Datasets}} & \multicolumn{6}{ c }{\textbf{Rich models}} \\
        \cline{2-7}
        & \multicolumn{2}{ c }{CRM} & \multicolumn{2}{ c }{SGRM} & \multicolumn{2}{ c }{GCRM} \\
        \cline{2-7}
        & 0.2 bpc & 0.4 bpc & 0.2 bpc & 0.4 bpc & 0.2 bpc & 0.4 bpc \\
        \hline
        \multicolumn{7}{ c }{\textbf{For HILL steganography}} \\
        \hline
        BOSS-PPG-LAN & \underline{0.6826} & \fbox{0.8061} & 0.68   & 0.8048 & 0.6743 & 0.8005 \\
        BOSS-PPG-CRP & \underline{0.9632} & 0.9952 & 0.9611 & 0.9947 & 0.9627 & \fbox{0.9954} \\
        BOSS-AHD-LAN & \underline{0.6817} & \fbox{0.8075} & 0.6813 & 0.8068 & 0.6775 & 0.8022 \\
        BOSS-AHD-CRP & \underline{0.9642} & 0.9942 & 0.9606 & \fbox{0.9957} & 0.9632 & 0.9951 \\
        \hline
        \multicolumn{7}{ c }{\textbf{For CMD-C-HILL steganography}} \\
        \hline
        BOSS-PPG-LAN & 0.6325 & \fbox{0.7548} & \underline{0.6333} & 0.7538 & 0.6265 & 0.748 \\
        BOSS-PPG-CRP & \underline{0.9337} & 0.9941 & 0.9329 & 0.9931 & 0.9324 & \fbox{0.9947} \\
        BOSS-AHD-LAN & \underline{0.6363} & \fbox{0.7558} & 0.6338 & 0.7528  & 0.6314 & 0.7477 \\
        BOSS-AHD-CRP & \underline{0.9357} & 0.9935 & 0.9317 & 0.9919 & 0.9331 & \fbox{0.9942} \\
        \Xhline{2\arrayrulewidth}
      \end{tabular}
    }}
  \label{tab:sel_competitor}
\end{table}
In prior works, color image steganalyzers were evaluated on different
scenarios and datasets. In order to select the most challenging
scenarios and competitors of our proposed WISERNet, we conducted a
preliminary experiment in which the detection performance of three
state-of-the-art rich models for color image steganalysis,
CRM~\cite{goljan_wifs_2014}, GCRM~\cite{abdulrahman_scn_2016}, and
SGRM~\cite{abdulrahman_ihmmsec_2016} were evaluated on four different
datasets. As shown in Tab.~\ref{tab:sel_competitor}, different
demosaicking algorithms~(PPG or AHD) had little effect on the
performance of the three rich models. The down-sampled datasets,
e.g. BOSS-PPG-LAN and BOSS-AHD-LAN, were the most challenging
scenarios for rich models. When evaluated on down-sampled datasets,
the performance of CRM was always the best. The performances of the
three rich models on BOSS-PPG-CRP and BOSS-AHD-CRP, the datasets
without down-sampling, were similar, and were all approaching to
100\%. Therefore for brevity, all the rest experiments reported in
this paper were only conducted on down-sampled datasets, and only CRM
was selected as the representative rich-model based competitor.

As for deep-learning steganalyzers, we selected Ye's
model~\cite{ye_tifs_2017}, and Xu's model \#2~\cite{xu_ihmmsec_2017}
as the representative competitors in the experiments. For Ye's model,
the bottom convolutional layer was equipped with channel-wise
convolution with learnable weights. To be fair, its
selection-channel-aware version was not included in the
experiments. For Xu's model \#2, please note that it is designed for
gray-scale JPEG image steganalysis. Therefore in the experiments we
adopted an alternative version of Xu's model \#2 in which the bottom
convolutional layer were with 30 fixed SRM kernels and input channel
concatenation. For a detailed discussion regarding to the
configurations of the bottom convolutional layer of the competing
deep-learning steganalyzers, please refer to
Sect.~\ref{sec:impact_of_comp}.

Our implementation of WISERNet was based on Caffe
toolbox~\cite{jia_acmmm_2014}. Unless otherwise specified, the model
magnification factor of WISERNet was fixed to $\bm{n}=9$. It was
trained using mini-batch stochastic gradient descent with ``inv''
learning rate starting from 0.001~(power: 0.75; gamma: 0.0001;
weight\_decay: 0.0005) and a momentum fixed to 0.9. The batch size in
the training procedure was 16 and the maximum number of iterations was
set to $30 \times 10^4$. The source codes and auxiliary materials are
available for download from
GitHub~\footnote{\url{https://github.com/tansq/WISERNet}}.

We used the same batch size and maximum number of iterations in the
training procedure of Ye's model and Xu's model \#2. The settings of
all other hyper-parameters for those two deep-learning based
competitors followed what reported in the original
papers~\cite{ye_tifs_2017,xu_ihmmsec_2017}. In every experiment,
6,000 cover-stego pairs were randomly selected for training. The
remaining 4,000 cover-stego pairs were for testing. All experiments
were repeated ten times, and the mean of predictive accuracies on
testing set over ten repetitions were reported. The experiments
involved two types of steganalyzers, the rich-model based and the
deep-learning based. For those rich-model based steganalyzers, FLD
ensemble classifier with default settings~\cite{kodovsky_tifs_2012}
was utilized. For those deep-learning steganalyzers, include our
proposed WISERNet, 1,000 cover-stego pairs were further randomly
picked out from training set for validation. In each experiment, the
model was validated and saved every $1 \times 10^4$ iterations. The
one with the best validation accuracy was evaluated on the
corresponding testing set.~\footnote{All of the deep-learning
  steganalyzers were trained and tested on a GPU cluster with 80
  NVIDIA$^{\textrm{\textregistered}}$
  Tesla$^{\textrm{\textregistered}}$ P100 GPU cards. The rich-model
  based steganalyzers were trained and tested on a CPU cluster with
  200 cores.}

\subsection{Comparison to state of the art}
\label{sec:comparison}

\begin{figure*}[!t]
  \centering
  \subfloat[]{
    \label{fig:full_comp_cmdchill_ppg_lan}
    \includegraphics[width=0.36\textwidth,keepaspectratio]{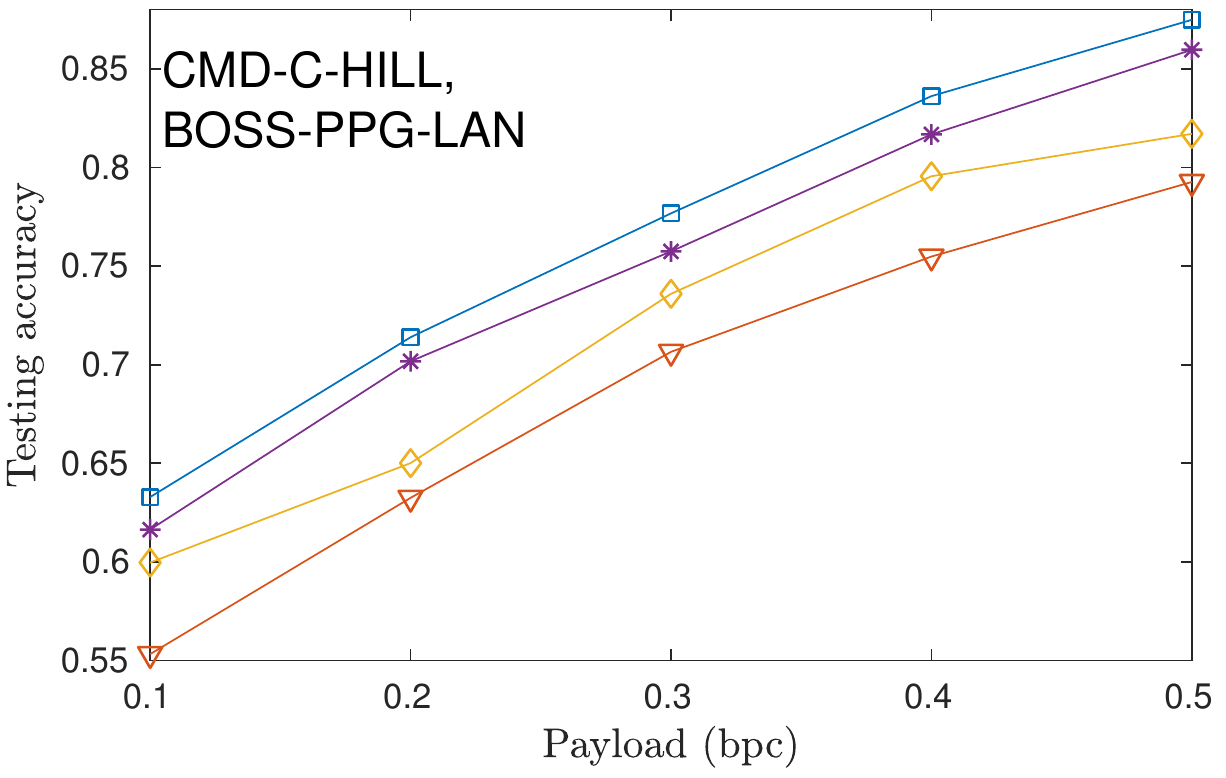} 
  }\hspace{.05\linewidth}
  \subfloat[]{
    \label{fig:full_comp_cmdcsuniward_ppg_lan}
    \includegraphics[width=0.36\textwidth,keepaspectratio]{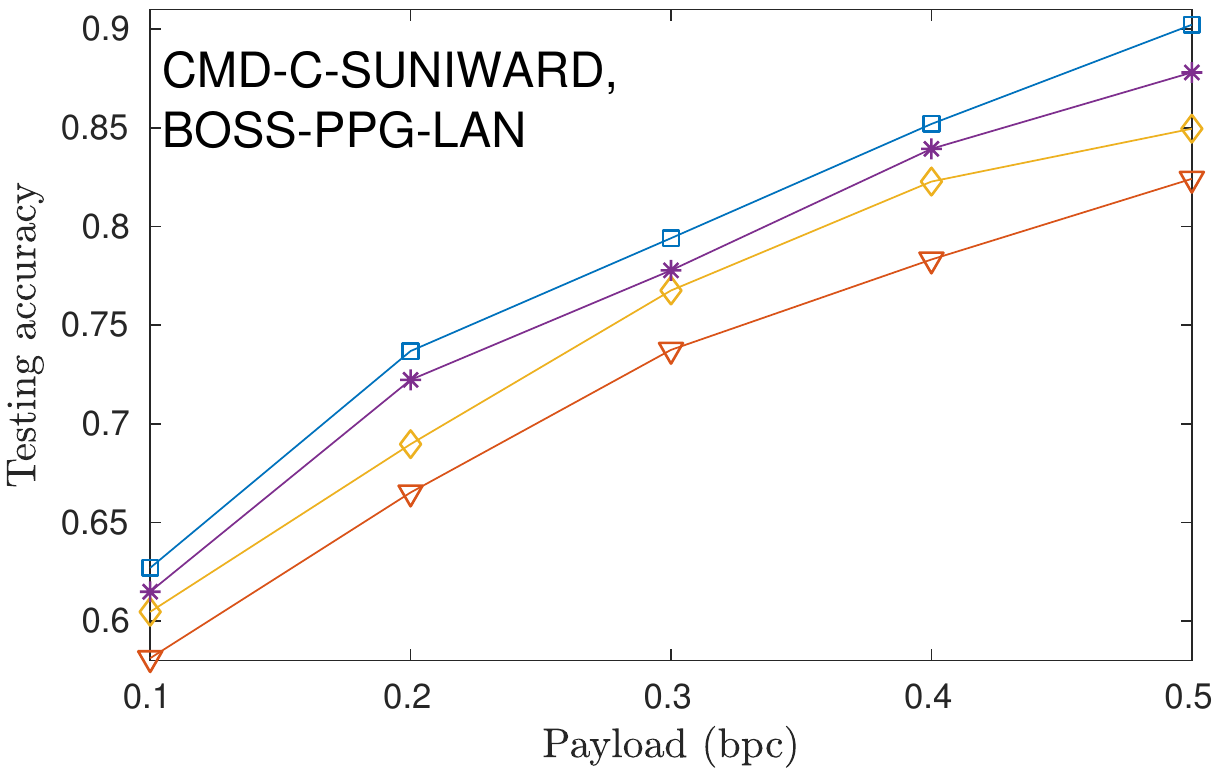}
  }\\
  \subfloat[]{
    \label{fig:full_comp_hill_ppg_lan}
    \includegraphics[width=0.36\textwidth,keepaspectratio]{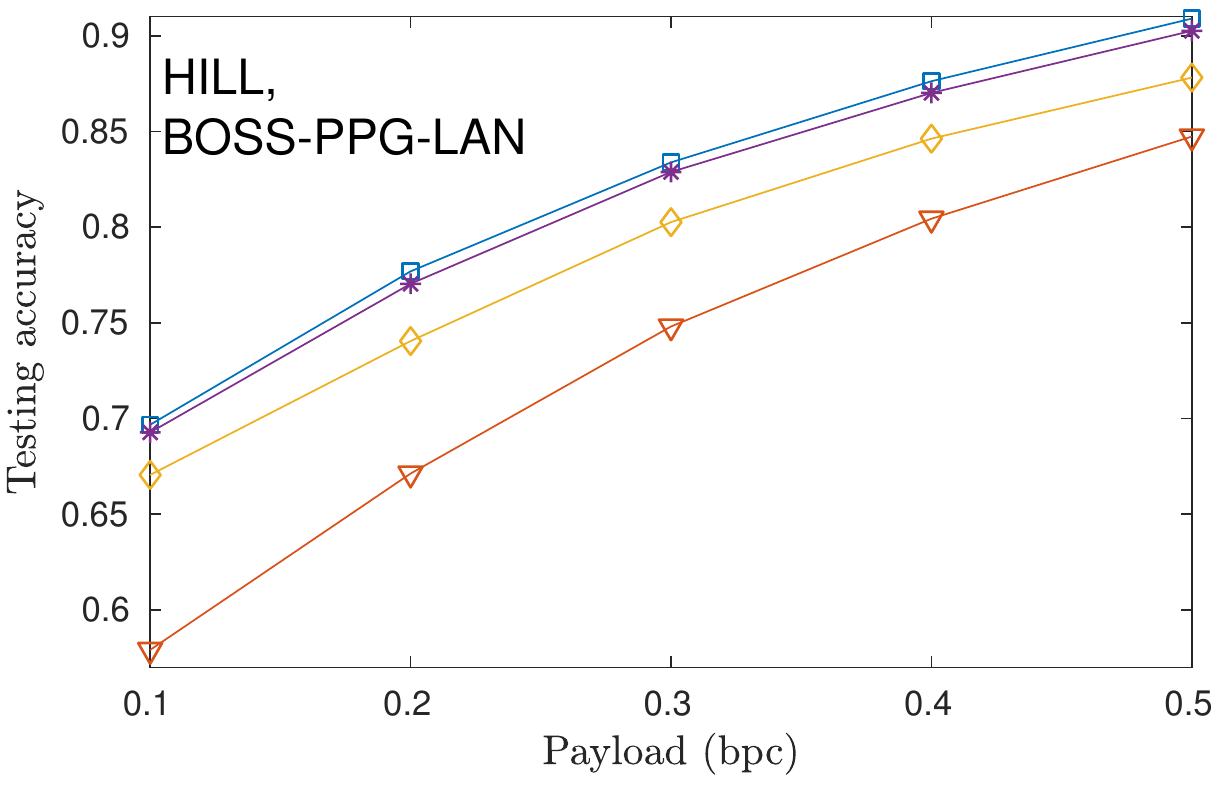} 
  }\hspace{.05\linewidth}
  \subfloat[]{
    \label{fig:full_comp_suniward_ppg_lan}
    \includegraphics[width=0.36\textwidth,keepaspectratio]{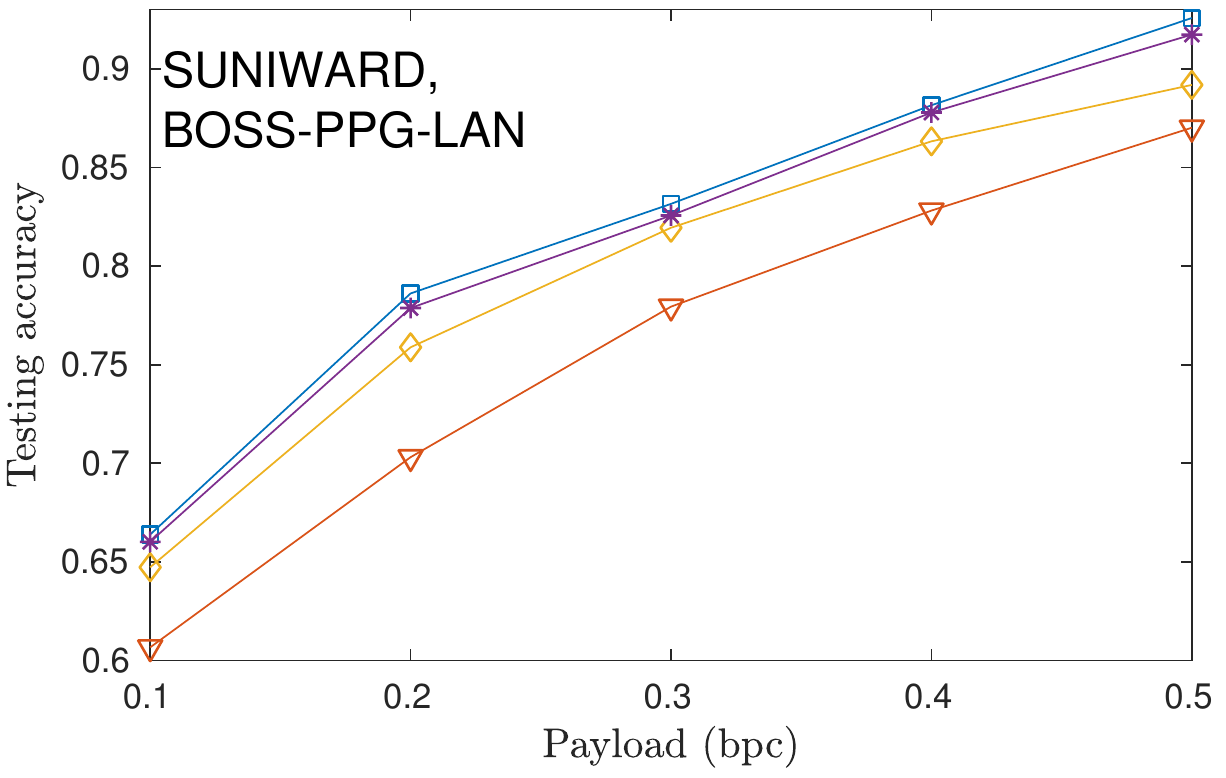}
  }\\
  \subfloat[]{
    \label{fig:full_comp_cmdchill_ahd_lan}
    \includegraphics[width=0.36\textwidth,keepaspectratio]{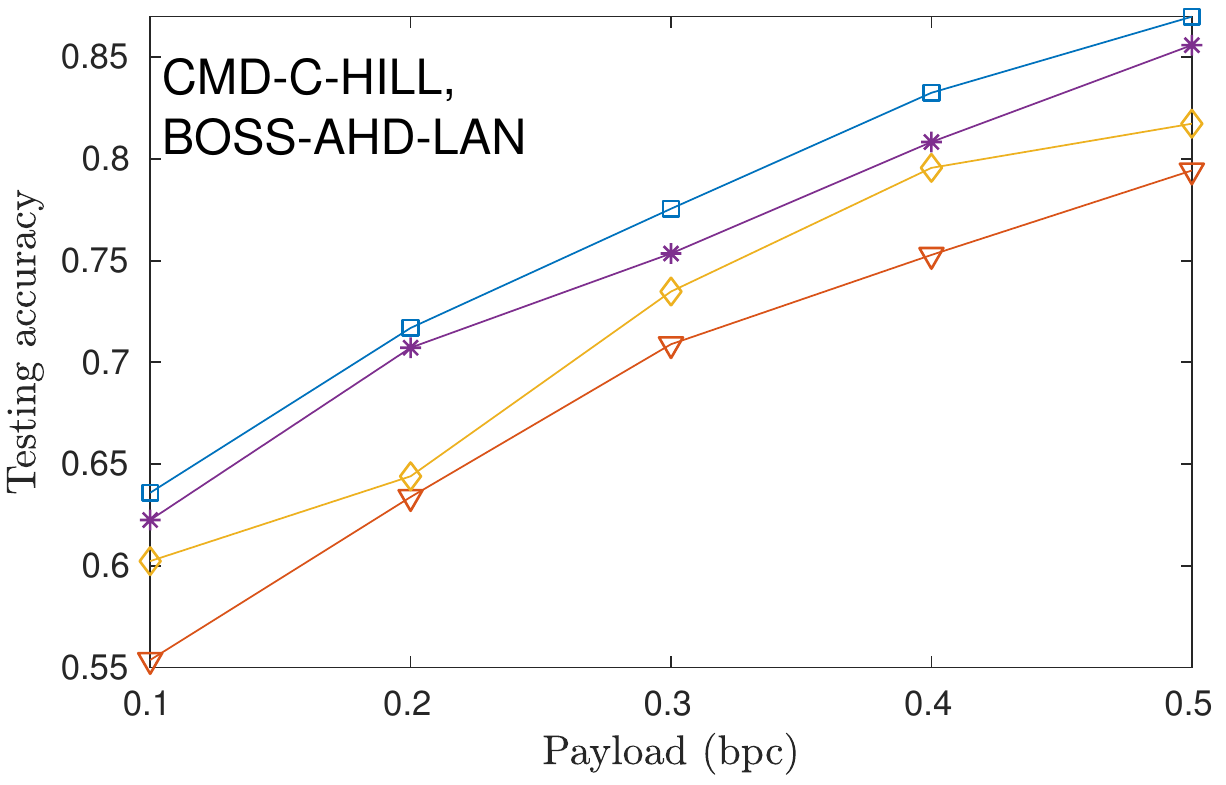} 
  }\hspace{.05\linewidth}
  \subfloat[]{
    \label{fig:full_comp_cmdcsuniward_ahd_lan} 
    \includegraphics[width=0.36\textwidth,keepaspectratio]{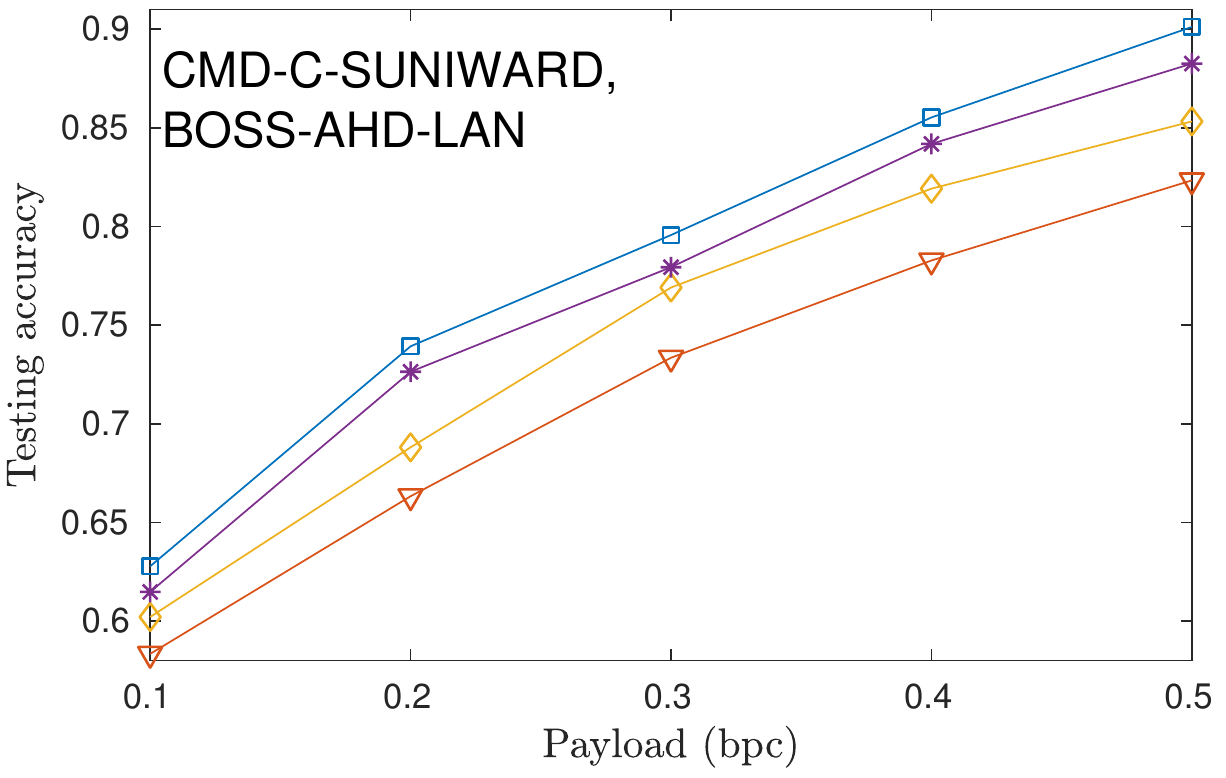}
  }\\
  \subfloat[]{
    \label{fig:full_comp_hill_ahd_lan}
    \includegraphics[width=0.36\textwidth,keepaspectratio]{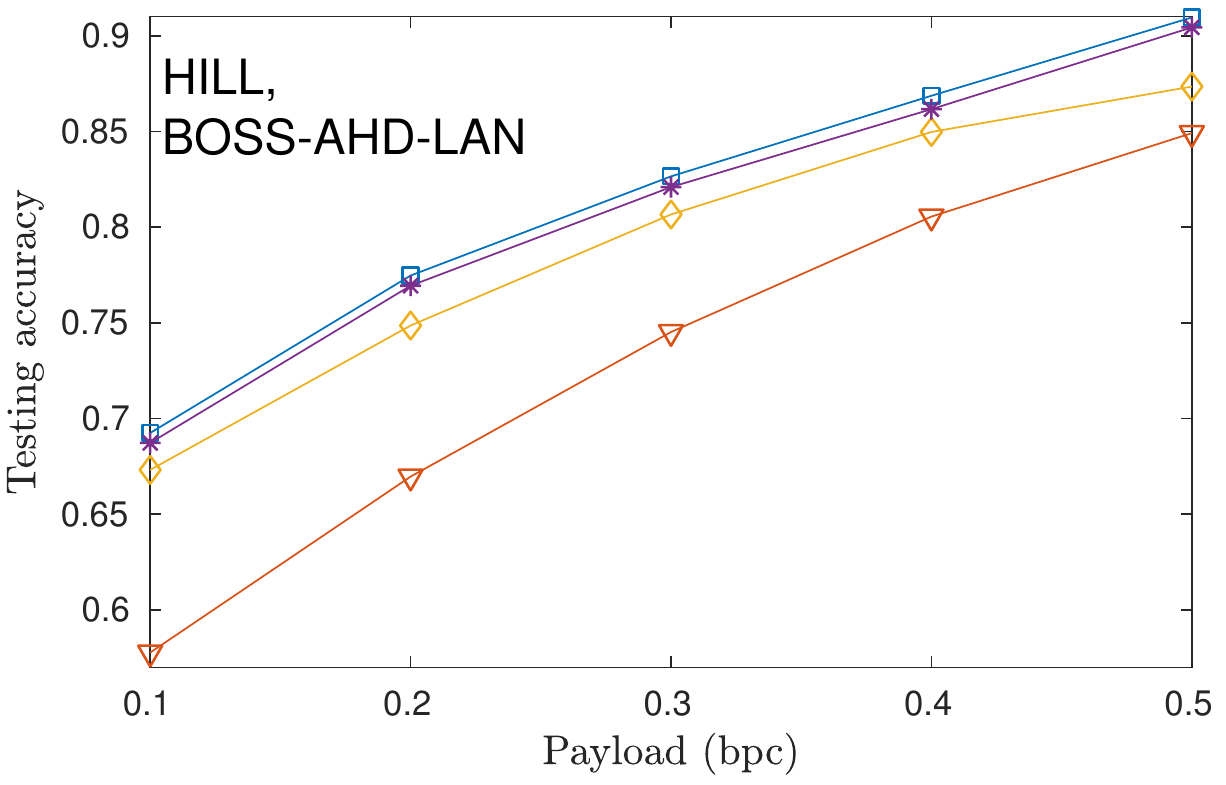} 
  }\hspace{.05\linewidth}
  \subfloat[]{
    \label{fig:full_comp_suniward_ahd_lan}
    \includegraphics[width=0.36\textwidth,keepaspectratio]{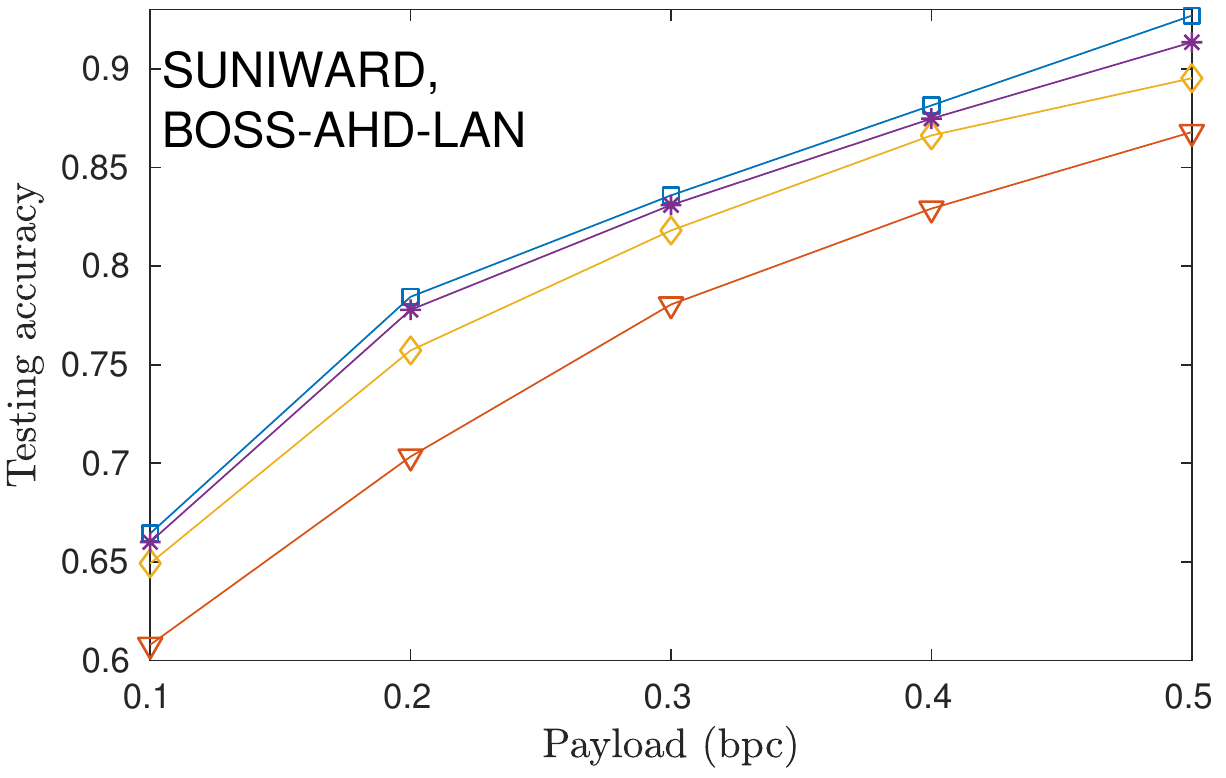}
  }\\
  \includegraphics[scale=0.8]{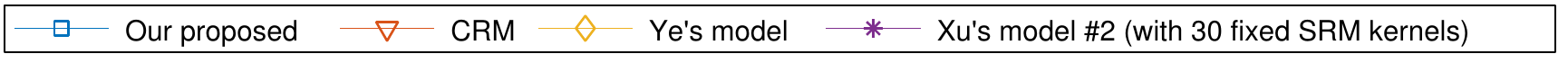} 
  \caption[]{Comparison of testing accuracy of our proposed WISERNet
    with state-of-the-art steganalyzers in the
    literature. \subref{fig:full_comp_cmdchill_ppg_lan} and
    \subref{fig:full_comp_cmdchill_ahd_lan} are the results for
    CMD-C-HILL; \subref{fig:full_comp_cmdcsuniward_ppg_lan} and
    \subref{fig:full_comp_cmdcsuniward_ahd_lan} are for
    CMD-C-SUNIWARD; \subref{fig:full_comp_hill_ppg_lan} and
    \subref{fig:full_comp_hill_ahd_lan} are the results for HILL;
    \subref{fig:full_comp_suniward_ppg_lan} and
    \subref{fig:full_comp_suniward_ahd_lan} are for SUNIWARD. The
    experiments for \subref{fig:full_comp_cmdchill_ppg_lan},
    \subref{fig:full_comp_cmdcsuniward_ppg_lan},
    \subref{fig:full_comp_hill_ppg_lan} and
    \subref{fig:full_comp_suniward_ppg_lan} were conducted on
    BOSS-PPG-LAN, while those for
    \subref{fig:full_comp_cmdchill_ahd_lan},
    \subref{fig:full_comp_cmdcsuniward_ahd_lan},
    \subref{fig:full_comp_hill_ahd_lan} and
    \subref{fig:full_comp_suniward_ahd_lan} were conducted on
    BOSS-AHD-LAN.}
  \label{fig:full_comp_state_of_the_art_diff_targets}
\end{figure*}

\begin{table*}[!t]
  \centering
  \caption[]{Full comparison of testing accuracy of our proposed
    WISERNet with state-of-the-art steganalyzers in the
    literature. CMD-C-HILL stego images with 0.2 bpc and 0.4 bpc were
    included. The results on BOSS-PPG-BIC, BOSS-PPG-BIL, BOSS-AHD-BIC
    and BOSS-AHD-BIL are given. The results on BOSS-PPG-LAN are also
    listed here for reference. The best results for 0.2 bpc are
    underlined, while the best results for 0.4 bpc are in framed
    boxes.}
  \resizebox{0.95\textwidth}{!}{%
    {\renewcommand{\arraystretch}{2}
      \begin{tabular}{ccccccccccccccccc}
        \Xhline{2\arrayrulewidth}
        \multirow{3}{*}{\large\textbf{Datasets}} & \multicolumn{2}{ c }{\multirow{2}{*}{\large \textbf{CRM}}} &
        \multicolumn{6}{ c }{\large \textbf{Ye's model}} &
        \multicolumn{6}{ c }{{\large \textbf{Xu's model \#2}}~(with 30 fixed SRM kernels)} &
        \multicolumn{2}{ c }{\multirow{2}{*}{\large \textbf{Our proposed}}} \\
        \cline{4-15}
        & & & 
        \multicolumn{2}{ c }{\makecell{Channel-wise \\ convolution}} &
        \multicolumn{2}{ c }{\makecell{Normal \\ convolution}} &
        \multicolumn{2}{ c }{\makecell{Input \\ concatenation}} &
        \multicolumn{2}{ c }{\makecell{Channel-wise \\ convolution}} &
        \multicolumn{2}{ c }{\makecell{Normal \\ convolution}} &
        \multicolumn{2}{ c }{\makecell{Input \\ concatenation}} & & \\
        \cline{2-17}
        & 0.2 bpc & 0.4 bpc &     
        0.2 bpc & 0.4 bpc & 0.2 bpc & 0.4 bpc & 0.2 bpc & 0.4 bpc & 
        0.2 bpc & 0.4 bpc & 0.2 bpc & 0.4 bpc & 0.2 bpc & 0.4 bpc & 
        0.2 bpc & 0.4 bpc \\    
        \hline
        BOSS-PPG-LAN & 0.6325 & 0.7548 & 0.6567 & 0.7965 & 0.6474 & 0.7568 & 0.6187 & 0.7169 & 0.6962 & 0.7941 & 0.6562 & 0.7763 & 0.7083 & 0.8167 & \underline{0.7139} & \fbox{0.8361} \\
        BOSS-PPG-BIC & 0.6551 & 0.7802 & 0.6741 & 0.8069 & 0.6589 & 0.7732 & 0.6332 & 0.7335 & 0.7124 & 0.8068 & 0.6611 & 0.7895 & 0.7294 & 0.8289 & \underline{0.7318} & \fbox{0.8435} \\
        BOSS-PPG-BIL & 0.7556 & 0.8717 & 0.7859 & 0.9019 & 0.7611 & 0.8721 & 0.7445 & 0.8343 & 0.7872 & 0.9045 & 0.7487 & 0.863 & 0.7904 & 0.9093 & \underline{0.8033} & \fbox{0.9169} \\
        BOSS-AHD-BIC & 0.6597 & 0.7832 & 0.6711 & 0.7991 & 0.6614 & 0.7728 & 0.6374 & 0.7355 & 0.7105 & 0.8141 & 0.6627 & 0.7922 & 0.7276 & 0.8198 & \underline{0.7369} & \fbox{0.8448} \\
        BOSS-AHD-BIL & 0.7578 & 0.8728 & 0.7804 & 0.9019 & 0.7622 & 0.8738 & 0.7376 & 0.837 & 0.7857 & 0.9067 & 0.7647 & 0.8593 & 0.7933 & 0.9061 & \underline{0.8022} & \fbox{0.9144} \\
        \Xhline{2\arrayrulewidth}
      \end{tabular}
    }}
  \label{tab:full_comp_state_of_the_art_diff_datasets}
\end{table*}

\begin{table*}[!t]
  \centering
  \caption[]{Comparison of testing accuracy of our proposed WISERNet
    with state-of-the-art steganalyzers in the literature under
    the scenario of mixed datasets. CMD-C-HILL stego images with
    0.2 bpc and 0.4 bpc were included. The best results for 0.2 bpc are
    underlined, while the best results for 0.4 bpc are in framed
    boxes. Abbreviations enclosed in braces indicates different
    involved datasets. For instance, BOSS-PPG-$\big\{$LAN, BIL,
    BIC$\big\}$ indicates that the mixed dataset is composed of
    BOSS-PPG-LAN, BOSS-PPG-BIL, and BOSS-PPG-BIC.}
  \resizebox{0.68\textwidth}{!}{%
    {\renewcommand{\arraystretch}{2}
      \begin{tabular}{ccccccccc}
        \Xhline{2\arrayrulewidth}
        \multirow{2}{*}{\large \textbf{Mixture of datasets}} & \multicolumn{2}{ c }{\large \textbf{CRM}} &
        \multicolumn{2}{ c }{{\makecell{{\large \textbf{Ye's model}} \\
        (with channel-wise convolution)}}} &
        \multicolumn{2}{ c }{{\makecell{{\large \textbf{Xu's model \#2}} \\
        (with 30 fixed SRM kernels, \\ and input concatenation)}}} &
        \multicolumn{2}{ c }{\large \textbf{Our proposed}} \\
        \cline{2-9}
        & 0.2 bpc & 0.4 bpc &     
        0.2 bpc & 0.4 bpc &     
        0.2 bpc & 0.4 bpc &     
        0.2 bpc & 0.4 bpc \\    
        \hline
        BOSS-PPG-$\big\{$LAN, BIL, BIC$\big\}$ & 0.6951 & 0.8136 & 0.7417 & 0.8379 & 0.7566 & 0.8497 & \underline{0.7658} & \fbox{0.8674} \\
        BOSS-AHD-$\big\{$LAN, BIL, BIC$\big\}$ & 0.6973 & 0.8193 & 0.7422 & 0.8402 & 0.7517 & 0.8516 & \underline{0.7649} & \fbox{0.8737} \\
        BOSS-$\big\{$PPG, AHD$\big\}$-LAN & 0.668 & 0.7878 & 0.6844 & 0.8031 & 0.7055 & 0.8118 & \underline{0.725} & \fbox{0.8411} \\
        Mixture of all six datasets & 0.6926 & 0.8101 & 0.7388 & 0.8401 & 0.7586 & 0.8516 & \underline{0.7633} & \fbox{0.8622} \\
        \Xhline{2\arrayrulewidth}
      \end{tabular} 
    }} 
  \label{tab:datasets_mixture}
\end{table*}

\begin{table*}[!t]
    \centering
    \caption[]{Comparison of number of parameters and computational
      complexity for our proposed WISERNet and other state-of-the-art
      deep-learning based steganalyzers. The computational
      complexity is measured in terms of FLOPs~(floating-point
      operations).}
   \label{tab:compare_model_parameters}
  \resizebox{0.85\textwidth}{!}{%
   {\renewcommand{\arraystretch}{1.4}
   \begin{tabular}{|c|c|c|c|c|}
       \hline
       & \textbf{Our proposed WISERNet} & \makecell{\textbf{Xu's model
                                          \#2} \\ \textbf{11-layer version} \\
        {\scriptsize (with 30 fixed SRM kernels,} \\ {\scriptsize and
       input concatenation)}} & \makecell{\textbf{Ye's model} \\
        {\scriptsize (with channel-wise convolution)}} & \makecell{\textbf{Xu's model \#2} \\
        {\scriptsize (with 30 fixed SRM kernels,} \\ {\scriptsize and input concatenation)}} \\
       \hline
       Parameters & $2.12 \times 10^6$ & $2.66 \times 10^6$ & $1.07 \times 10^6$ & $4.87 \times 10^6$ \\
       \hline
       FLOPs & $4.11 \times 10^9$ & $1.08 \times 10^{10}$ & $5.76 \times 10^9$ & $1.68 \times 10^{10}$ \\
       \hline
     \end{tabular}
   }
 }
\end{table*}

Firstly, we report the results of the experiments conducted on
BOSS-PPG-LAN and BOSS-AHD-LAN. From
Fig.~\ref{fig:full_comp_state_of_the_art_diff_targets} we can see
different demosaicking algorithms~(PPG or AHD) had little impact on
the performance of color image steganalyzers. All of the three
deep-learning based models could obtain significant performance
improvement compared with CRM, the rich-model based steganalyzer. As
for the three deep-learning steganalyzers themselves, the performance
of Xu's model \#2~(with 30 fixed SRM kernels) was always better than
that of Ye's model, which is reasonable since compared with Ye's
model, Xu's model \#2 is much deeper and more complicated. However,
our proposed WISERNet performed even better than Xu's model \#2
although it is a relatively shallow and small model.

From the perspective of color image steganography, it is no doubt that
CMD-C better resisted color image steganalysis.  However, the
superiority of WISERNet was also more obvious when used to attack
CMD-C steganography. For CMD-C-HILL with 0.4 bpc on BOSS-AHD-LAN
dataset, WISERNet could further increase detection accuracy by as
large as 4\% on the basis of Xu's model \#2~(with 30 fixed SRM
kernels). The more obvious performance improvement for CMD-C implies
that compared with Xu's model \#2, the specific shallow-and-wide,
separate-then-reunion structure of WISERNet can better utilize the
intrinsic statistical characteristics among color bands to attack
CMD-C.

For the sake of completeness, for CMD-C-HILL, we give a full
comparison of testing accuracy of our proposed WISERNet with CRM, Ye's
model and Xu's model \#2 in
Tab.~\ref{tab:full_comp_state_of_the_art_diff_datasets}. The results
for two representative payloads, 0.2 bpc and 0.4 bpc are given. The
impacts of another two down-sampling algorithms, ``Bicubic'', and
``Bilinear'' are inspected. Therefore the results on BOSS-PPG-BIC,
BOSS-PPG-BIL, BOSS-AHD-BIC and BOSS-AHD-BIL are listed. From
Tab.~\ref{tab:full_comp_state_of_the_art_diff_datasets} we can see
different down-sampling algorithms had huge impact on detection
performance of the steganalyzers. The steganalyzers always achieved
better performances on the images generated with ``Bilinear''
down-sampling algorithms. For Ye's model, channel-wise convolution in
the bottom convolutional layer was always the best choice. However for
Xu's model \#2, with input band concatenation was the best
choice. Although detection performance of WISERNet was also affected
by different down-sampling algorithms, in all scenarios it performed
the best and possessed a clear margin of superiority.

In Tab.~\ref{tab:datasets_mixture}, we also give a comparison of
testing accuracy of the four steganalyzers under the scenario of
mixed datasets for CMD-C-HILL. In each experiment, we used the same
pseudo-random number series to  split every involved dataset
into training set, validation set~(for deep-learning steganalyzers), and
testing set. The corresponding subsets were merged together to form
the mixed training set, validation set and testing set,
respectively. Therefore we guaranteed that all the cover-stego pairs
generated from the same raw image can only be included in one of the
mixed subsets, e.g. in the training set. Firstly we fixed the
demosaicking algorithm, and mixed the datasets with three different
down-sampling options, namely with ``Lanczos'' kernel of ImageMagick,
``Bicubic'' and ``Bilinear'' kernels of
Matlab$^{\textrm{\textregistered}}$. Then we fixed the down-sampling
option to ``Lanczos'' and mixed the datasets with PPG and AHD
demosaicking algorithms. Finally we mixed the datasets with two
demosaick options and three down-sampling options. From
Tab.~\ref{tab:datasets_mixture} we can see the effects of different
demosaicking options and down-sampling options on detection
performance of the investigated steganalyzers are similar. But
it is amazing that in all of those scenarios our proposed WISERNet
always performed the best.

\begin{table*}[!t]
  \centering
  \caption[]{Comparison of testing accuracy of our proposed WISERNet
    with the 11-layer version of Xu's model \#2~(with 30 fixed SRM
    kernels). The experiments were conducted on BOSS-PPG-LAN. The
    terms in parentheses with preceding \textuparrow\ denote accuracy
    increment of our proposed WISERNet compared to the 11-layer
    version of Xu's model \#2.}
  \resizebox{0.85\textwidth}{!}{%
    {\renewcommand{\arraystretch}{1.2}
      \begin{tabular}{cccccc}
        \hline
        \multirow{2}{*}{\textbf{Steganalyzers}} & \multicolumn{5}{ c }{\textbf{Payload~(bpc)}} \\ \cline{2-6}
        & 0.1 & 0.2 & 0.3 & 0.4 & 0.5 \\
        \hline
        Our proposed WISERNet & 0.6329~\textbf{(\textuparrow 0.0321)}& 0.7139~\textbf{(\textuparrow 0.0364)} & 0.7767~\textbf{(\textuparrow 0.0376)} & 0.8361~\textbf{(\textuparrow 0.0412)} & 0.8748~\textbf{(\textuparrow 0.0433)} \\ 
        \hline
        \makecell{11-layer version of \\ Xu's model \#2} & 0.6008 & 0.6775 & 0.7391 & 0.7949 & 0.8315 \\
        \hline
      \end{tabular}      
    }}
  \label{tab:performance_cmdc_11_layer_xunet}
\end{table*}

The above experimental results indicates that our proposed WISERNet is
with superior performance especially when used to attack CMD-C, the
latest steganographic algorithm purposely for color images. The
superiority of WISERNet is obvious even compared with the most
advanced and complicated deep-learning steganalyzer, namely Xu's model
\#2. Please note that the detection performance gain is not achieved
with deeper, larger or more complicated deep-learning structure. As
shown in Tab.~\ref{tab:compare_model_parameters}, WISERNet is only
with less than half parameters and about a quarter computational
complexity compared with Xu's model \#2~(with fixed 30 SRM kernels),
the second best performing deep-learning steganalyzer after
WISERNet. If we would like to make a fairer comparison, the 11-layer
version of Xu's model \#2 is a suitable rival, which was also
investigated in \cite{xu_ihmmsec_2017}. The 11-layer version of Xu's
model \#2 possesses slightly more parameters and more than double
computational complexity compared with our proposed WISERNet. But as
shown in Tab.~\ref{tab:performance_cmdc_11_layer_xunet}, the
superiority of WISERNet is obvious. For instance, when the payload of
the target CMD-C is 0.4bpc, WISERNet can surpass the 11-layer version
of Xu's model \#2 by more than four percent.

\subsection{Impact of different components for WISERNet and its
  deep-learning based competitors}
\label{sec:impact_of_comp}

\setul{0.5ex}{0.3ex}
\begin{table*}[!t]
  \centering
  \caption[]{Impact of different configurations of the bottom
    convolutional layer. The experiments were conducted on
    BOSS-PPG-LAN. Only CMD-C-HILL stego images with 0.4 bpc were
    included. The best result in every column is underlined. The underline
    in bold highlights the best result among them.}
  \resizebox{0.65\textwidth}{!}{%
    {\renewcommand{\arraystretch}{1.3}
      \begin{tabular}{ccc|ccc}
        \Xhline{2\arrayrulewidth}
        \multirow{4}{6em}{\textbf{\makecell{How to \\ convolve}}} & 
        \multirow{4}{*}{\textbf{Learnable}} & 
        \multirow{4}{*}{\makecell{\textbf{Our} \\ \textbf{proposed} \\ \textbf{network}}} & 
        \multicolumn{3}{ c }{\textbf{Compared deep-learning steganalyzers}} \\
        \cline{4-6}
        & & &
        \multirow{3}{*}{\makecell{\textbf{Ye's} \\ \textbf{model}}} &
        \multirow{3}{*}{\makecell{\textbf{Xu's model \#2} \\ with 16 \\ DCT kernels}} &
        \multirow{3}{*}{\makecell{\textbf{Xu's model \#2} \\ with 30 \\ SRM kernels}} \\
       & & & & & \\
       & & & & & \\
       \hline
       \multirow{2}{6em}{\makecell{Channel-wise \\ convolution}} & \cmark & \ul{0.8361} & \underline{0.7965} & 0.7244 & 0.7941 \\
       & \xmark & 0.8232 & 0.7895 & 0.7269 & 0.7951 \\
       \hline
       \multirow{2}{6em}{\makecell{Normal \\ convolution}} & \cmark & 0.7268 & 0.7608 & 0.6916 & 0.7725 \\
       & \xmark & 0.7257 & 0.7566 & 0.7085 & 0.7763 \\
       \hline
       \multirow{2}{6em}{\makecell{Input \\ concatenation}} & \cmark & 0.7259 & 0.7319 & 0.7672 & 0.7987 \\
       & \xmark & 0.7270 & 0.7268 & \underline{0.7916} & \underline{0.8167} \\
        \hline
       \multirow{2}{6.5em}{\makecell{Cross-band \\ interleave}} & \cmark & 0.7063 & 0.6935 & 0.7051 & 0.7122 \\
       & \xmark & 0.6968 & 0.6848 & 0.6936 & 0.7041 \\
        \Xhline{2\arrayrulewidth}
      \end{tabular}
    }}
  \label{tab:bottom_layer}
\end{table*}

\begin{table*}[!t]
  \centering
  \caption[]{Impact of different  model magnification factor $\bm{n}$. The experiments were conducted on
    BOSS-PPG-LAN. Only CMD-C-HILL stego images with 0.4 bpc were
    included. The best result is in framed box.}
  \resizebox{0.75\textwidth}{!}{%
    {\renewcommand{\arraystretch}{1.5}
      \begin{tabular}{cccccccccc}
        \Xhline{2\arrayrulewidth}
        \multicolumn{10}{ c }{\textbf{Model magnification factor} {$\bm{n}$}}\\
        \hline
        1 & 2 & 3 & 4 & 5 & 6 & 7 & 8 & 9 & 10 \\
        \hline
        0.7375 & 0.8037 & 0.8208 & 0.8214 & 0.8227 & 0.8262 & 0.8313 & 0.8319 & \fbox{0.8361} & 0.8321 \\
        \Xhline{2\arrayrulewidth}
      \end{tabular}
    }}
  \label{tab:model_magnification}
\end{table*}

In Tab.~\ref{tab:bottom_layer} we further compare the impact of
different configurations of the bottom convolutional layer on
detection performances of our proposed WISERNet, and another two
deep-learning based competitors. From Tab.~\ref{tab:bottom_layer} we
can see for our proposed WISERNet, channel-wise convolution with
learnable weights always achieved the best performance.  The results
also show that channel-wise convolution with learnable weights was
more suitable for Ye's model. Besides, we can clearly observe that
with cross-band interleave pre-processing strategy, all deep-learning
based steganalyzers suffer severe performance degradation.

Xu's model \#2 is the deepest and the most advanced deep-learning
steganalyzer among the competitors. However it is designed for
gray-scale JPEG image steganalysis, and the 16 $4 \times 4$ DCT
high-pass filters in the bottom convolutional layer adopted in its
original version might not suitable for spatial-domain color image
steganalysis. Tab.~\ref{tab:bottom_layer} clearly shows that the
alternative version of Xu's model \#2~(in which the 16 DCT filters are
replaced with 30 SRM kernels) performed better than the original
version. It is noteworthy that with fixed kernel weights, input
band concatenation before the bottom convolutional layer is the
best option for Xu's model \#2. It may be attributed to the very deep
structure of Xu's model \#2, which makes the convolution kernels in
the top layers finally have the opportunity to perceive the
cross-band correlation in pixels.

Tab.~\ref{tab:model_magnification} shows the impact of different model
magnification factor $\bm{n}$ on our proposed WISERNet. As shown in
Tab.~\ref{tab:model_magnification}, starting from $\bm{n}=1$,
detection performance was steadily promoted along with increasing
$\bm{n}$, and reached its maximum with $\bm{n}=9$. Therefore, $\bm{n}$
was fixed to $9$ in our experiments.

\subsection{Imapct of learnable bottom kernels}
\label{sec:impact_learnable_bottom_kernels}

\begin{table*}[!t]
    \centering
    \caption[]{ $\overline{C_W}$, $\mathring{\overline{|S|}}_C$ and
      $\mathring{\overline{|S|}}_S$ versus increasing iterations of our proposed
      WISERNet. The experiment was conducted on
      BOSS-PPG-LAN. CMD-C-HILL stego images with 0.4 bpc were
      included.}
   \label{tab:avg_corr_vs_iterations}
  \resizebox{0.6\textwidth}{!}{%
   {\renewcommand{\arraystretch}{1.4}
     \begin{tabular}{|c|c|c|c||c|c|c|c|}
       \hline
       \textbf{Iterations} & $\overline{C_W}$ & $\mathring{\overline{|S|}}_C$ & $\mathring{\overline{|S|}}_S$ & \textbf{Iterations} & $\overline{C_W}$ & $\mathring{\overline{|S|}}_C$ & $\mathring{\overline{|S|}}_S$  \\
       \hline
       $0 \times 10^4$ & 1      & 0.9888 & 0.9888  & $6 \times 10^4$ & 0.9877  & 0.9886 & 0.9886 \\
       \hline                                                                                   
       $1 \times 10^4$ & 0.9985 & 0.9888 & 0.9888  & $10 \times 10^4$ & 0.9876 & 0.9884 & 0.9884 \\
       \hline                                                                                   
       $2 \times 10^4$ & 0.9978 & 0.9886 & 0.9886  & $15 \times 10^4$ & 0.9814 & 0.9881 & 0.9881 \\
       \hline                                                                                   
       $3 \times 10^4$ & 0.9978 & 0.9886 & 0.9886  & $20 \times 10^4$ & 0.9753 & 0.9889 & 0.9889 \\
       \hline                                                                                   
       $4 \times 10^4$ & 0.9975 & 0.9890 & 0.9890  & $25 \times 10^4$ & 0.9795 & 0.9886 & 0.9886 \\
       \hline                                                                                   
       $5 \times 10^4$ & 0.9967 & 0.9891 & 0.9891  & $30 \times 10^4$ & 0.9848 & 0.9887 & 0.9887 \\
       \hline
     \end{tabular}
   }}
\end{table*}

Firstly, we try to give an explanation why making the weights of the
bottom convolution kernels of WISERNet learnable can further improve
detection performance.

We agree with the opinion that the main purpose of the convolutions in
the bottom convolutional layer is to suppress image contents and
retain stego noises at the same
time~\cite{tan_apsipa_2014,qian_spie_2015,xu_spl_2016,ye_tifs_2017}. However,
as pointed out in \cite{zeng_tifs_2018}, optimization of the bottom
convolution kernels in favor of the extraction of stego noises is hard
to achieve with gradient-descent based learning. In fact, as mentioned
in Sect.~\ref{sec:preliminaries}, fixed high-pass filters in the
bottom convolutional layer of most existing deep-learning
steganalyzers also provide evidences to support our argument.

Since the proposal of rich models~\cite{fridrich_tifs_2012}, it is
well accepted that model diversity is crucial to the performance of
steganalyzers. Therefore we believe the performance improvement of
WISERNet can be attributed to the further model diversity brought by
continuous learning and optimizing of the kernels in the bottom
convolutional layer of WISERNet, although learnable bottom kernels
cannot help boost SNR. Refer to Sect.~\ref{sec:rationale}, the bottom
convolution kernels of WISERNet can be divided into triples
$\big\{\textrm{\textbf{W}}^{1}_{1k}, \textrm{\textbf{W}}^{1}_{2k},
\textrm{\textbf{W}}^{1}_{3k}\big\}, 1 \le k \le 30$.  Denote the
average of pair-wise correlation between weights in the triples as:
\begin{equation}
  \label{eq:avg_corr_w}
  \overline{C_W}=\frac{\sum_{k=1}^{30}\sum_{1 \le i < j \le 3}\textrm{Corr}(\textrm{\textbf{W}}^{1}_{ik},\textrm{\textbf{W}}^{1}_{jk})}{90}
\end{equation}
$\overline{C_W}$ can be used to measure the diversity of the bottom
kernels.  Initially, we set
$\textrm{\textbf{W}}^{1}_{1k}=\textrm{\textbf{W}}^{1}_{2k}=\textrm{\textbf{W}}^{1}_{3k}=\widetilde{\textrm{\textbf{W}}}_k$,
where $\widetilde{\textrm{\textbf{W}}}_k$ is one of the SRM high-pass
filters. Therefore initially $\overline{C_W}=1$, and it decreases
along with increasing diversity of the bottom kernels.  In
Tab.~\ref{tab:avg_corr_vs_iterations}, we give a demonstration. In one
training procedure of our proposed WISERNet which aimed at attacking
CMD-C-HILL stego images with 0.4 bpc, we could clearly inspect
$\overline{C_W}$ steadily decreased with increasing training
iterations, which indicates diversity of the bottom kernels increased
along with increasing training iterations.  $\overline{C_W}$ reached
its minimum at around $20 \times 10^4$ iterations. It is interesting
that WISERNet also achieved its best validation accuracy at
$20 \times 10^4$ iterations, which implies that the performance of
WISERNet was relevant with diversity of the bottom kernels.  However,
please note that initialized with high-pass filters, the bottom
kernels of WISERNet eventually cannot exhibit large diversity even
with enormous iterations. In our extensive experiments, we have never
observed $\overline{C_W}$ was reduced to below 0.9 even after
$100 \times 10^4$ iterations. At last, one more remarkable thing is
that Xu's model \#2, the one with much deeper structure could not gain
better performance with learnable bottom kernels. Therefore we believe
that the wide and shallow structure of WISERNet might be the
determining factor of beneficial learnable bottom convolution
kernels.

As mentioned in Sect.~\ref{sec:rationale}, our basic assumption is
that
$\textrm{E}(\textrm{\textbf{C}}_{1} \ast \textrm{\textbf{W}}^{1}_{1k})
\approx \textrm{E}(\textrm{\textbf{C}}_{2} \ast
\textrm{\textbf{W}}^{1}_{2k}) \approx
\textrm{E}(\textrm{\textbf{C}}_{3} \ast
\textrm{\textbf{W}}^{1}_{3k})$.  One interesting question arises: Is
this assumption still holds with more and more diverse bottom kernels?
For a given image, denote
$\textrm{E}(\textrm{\textbf{C}}_{i} \ast
\textrm{\textbf{W}}^{1}_{ik})=\mu_{ik}, 1 \le i \le 3$.  Let
$\boldsymbol{\mu}_k=(\mu_{1k},\mu_{2k},\mu_{3k})$,
$\boldsymbol{1}=(1,1,1)$, and $\theta_k$ be the angle between vector
$\boldsymbol{\mu}_k$ and vector $\boldsymbol{1}$. The \textit{cosine
  similarity}~\cite{wiki_cos_sim} between $\boldsymbol{\mu}_k$ and
$\boldsymbol{1}$ is defined as:
\begin{equation}
  \label{eq:measure_equal_mu}
  S_k=\cos(\theta_k)=\frac{\mu_{1k}+\mu_{2k}+\mu_{3k}}{\sqrt{3}\cdot \sqrt{\mu^2_{1k}+\mu^2_{2k}+\mu^2_{3k}}}
\end{equation}
However, since $\textrm{\textbf{W}}^{1}_{ik}, 1 \le i \le 3$ are
high-pass filters, $\mu_{ik}, 1 \le i \le 3$ may be less than zero. As
a result, $S_k$ ranges from $-1$ to $1$. For simplicity we use $|S_k|$ to
measure the similarity among $\mu_{ik}, 1 \le i \le 3$. The more
similar $\mu_{ik}, 1 \le i \le 3$ are, the more parallel $\boldsymbol{\mu}_k$ and
$\boldsymbol{1}$ tend to be , the more $|S_k|$ is close to 1.

Fed with one image, let $\overline{|S|}$ denote the average of
$|S_k|, 1 \le k \le 30$. In Tab.~\ref{tab:avg_corr_vs_iterations}, we
also analyzed the 1,000 cover-stego pairs in the validation set of
BOSS-PPG-LAN. Denote the average of $\overline{|S|}$ of the 1,000
cover images as $\mathring{\overline{|S|}}_C$, and that of the 1,000
stego images as $\mathring{\overline{|S|}}_S$. From
Tab.~\ref{tab:avg_corr_vs_iterations} we can see
$\mathring{\overline{|S|}}_C$ was always equal to
$\mathring{\overline{|S|}}_S$ which means that the impact of stego
noises was negligible. Though $\overline{C_W}$ slowly but steadily
decreased with increasing training iterations, both
$\mathring{\overline{|S|}}_C$ and $\mathring{\overline{|S|}}_S$ kept
close to 1. Therefore the experimental evidence indicated that for our
proposed WISERNet, the means of the corresponding output channels are 
still nearly equivalent even with more and more diverse bottom kernels
during the training procedure. As a result, our basic assumption in
Sect.~\ref{sec:rationale} holds.

\begin{table*}[!t]
  \centering
  \caption[]{Comparison of testing accuracy when the targets
    are CMD-C-HILL and its alternative version in which the correlation
    among three color bands is disabled. The experiments were
    conducted on BOSS-PPG-LAN. The terms in parentheses with preceding
    \textuparrow\ denote accuracy increment of our proposed WISERNet
    compared to Xu's model \#2~(with 30 fixed SRM kernels).}
  \resizebox{0.85\textwidth}{!}{%
    {\renewcommand{\arraystretch}{1.2}
      \begin{tabular}{cccccc}
        \hline
        \multirow{2}{*}{\textbf{Steganalyzers}} & \multicolumn{5}{ c }{\textbf{Payload~(bpc)}} \\ \cline{2-6}
        & 0.1 & 0.2 & 0.3 & 0.4 & 0.5 \\
        \hline
        \multicolumn{6}{ c }{\textbf{For CMD-C-HILL}} \\
        \hline
        Our proposed WISERNet & 0.6329~\textbf{(\textuparrow 0.0165)}& 0.7139~\textbf{(\textuparrow 0.0122)} & 0.7767~\textbf{(\textuparrow 0.0193)} & 0.8361~\textbf{(\textuparrow 0.0194)} & 0.8748~\textbf{(\textuparrow 0.0152)} \\ 
        \hline
        \makecell{Xu's model \#2 \\(with 30 fixed SRM kernels)} & 0.6164 & 0.7017 & 0.7574 & 0.8167 & 0.8596 \\
        \hline
        \multicolumn{6}{ c }{\makecell{\textbf{For the alternative version of
        CMD-C-HILL} \\(with the correlation
    among three color bands disabled)}} \\
        \hline
        Our proposed WISERNet & 0.6509~\textbf{(\textuparrow 0.0172)}& 0.7584~\textbf{(\textuparrow 0.0242)} & 0.7985~\textbf{(\textuparrow 0.0247)} & 0.8598~\textbf{(\textuparrow 0.0266)} & 0.8934~\textbf{(\textuparrow 0.0177)} \\ 
        \hline
        \makecell{Xu's model \#2 \\(with 30 fixed SRM kernels)} & 0.6337 & 0.7342 & 0.7738 & 0.8332 & 0.8757 \\
        \hline
      \end{tabular}      
    }}
  \label{tab:performance_cmdc_disable_inter_band_corr}
\end{table*}

\subsection{Impact of the correlations among color bands of the
  targets}
\label{sec:impact_corr_bands_of_targets}

Our major target, CMD-C is a non-additive embedding distortion
minimizing framework which can preserve not only the correlation
within each color band, but also the correlations among three color
bands. It is interesting to observe how the correlations among color
bands of CMD-C stego images affect the performance of our proposed
WISERNet. We can disable the inter-band correlations in CMD-C stego
images via removing elements from other bands in the calculation of
the costs~\cite{tang_spl_2016}. As shown in
Tab.~\ref{tab:performance_cmdc_disable_inter_band_corr}, when the
target is the alternative version of CMD-C in which the correlation
among three color bands is disabled, our proposed WISERNet achieves
better performance. Furthermore, the gap between Xu's model \#2 and
WISERNet is wider than when the target is the original CMD--HILL. The
wider gap implies that inter-band correlation of stego noises
introduced by original CMD-C~(even weak) do help it better resist the
channel-wise convolution in the bottom ``Separate'' stage of
WISERNet. Our proposed wide-and-shallow, separate-then-reunion network
structure shows even greater advantage when used to attack CMD-C
without the correlation among three color bands.

\subsection{Performance on large-scale dataset under cover-source
  mismatching scenarios}
\label{sec:performance_another_dataset}

\begin{figure*}[!t]
  \centering
  \subfloat[]{
    \label{fig:full_comp_cover_source_mismatching_b}
    \includegraphics[width=0.36\textwidth,keepaspectratio]{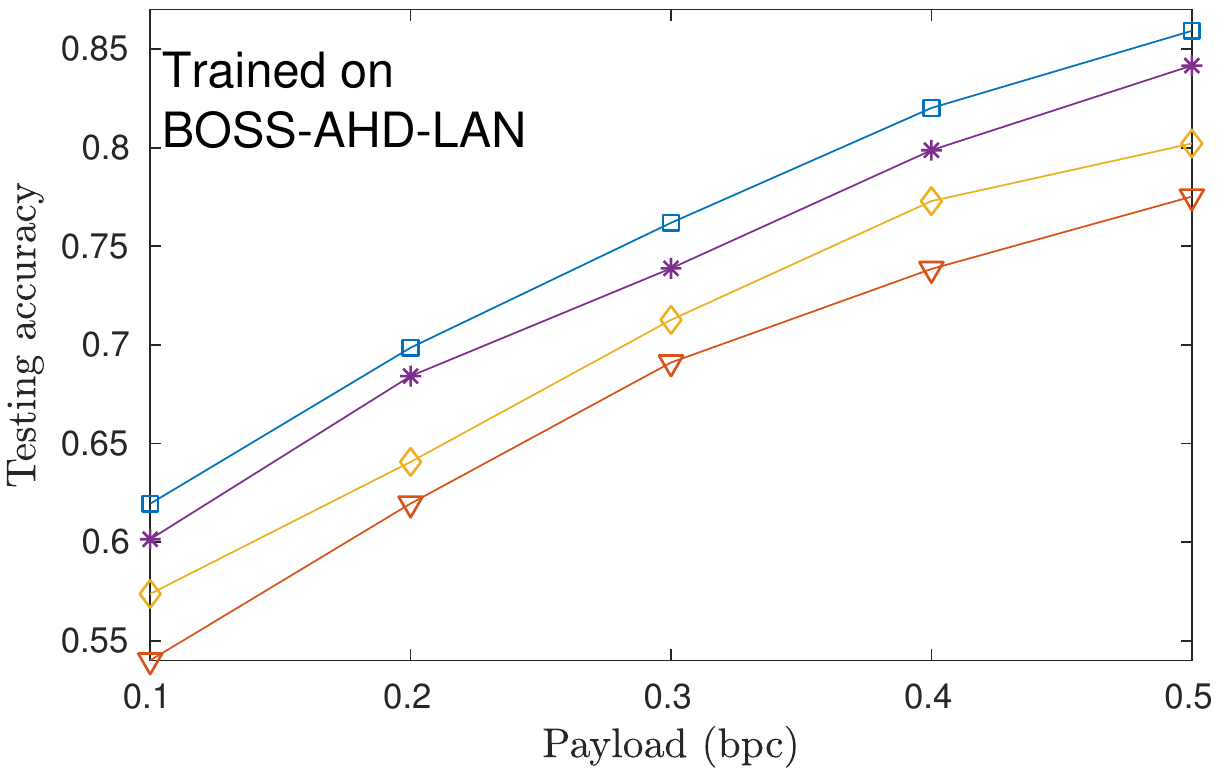}
  }\hspace{.05\linewidth}
  \subfloat[]{
    \label{fig:full_comp_cover_source_mismatching_c}
    \includegraphics[width=0.36\textwidth,keepaspectratio]{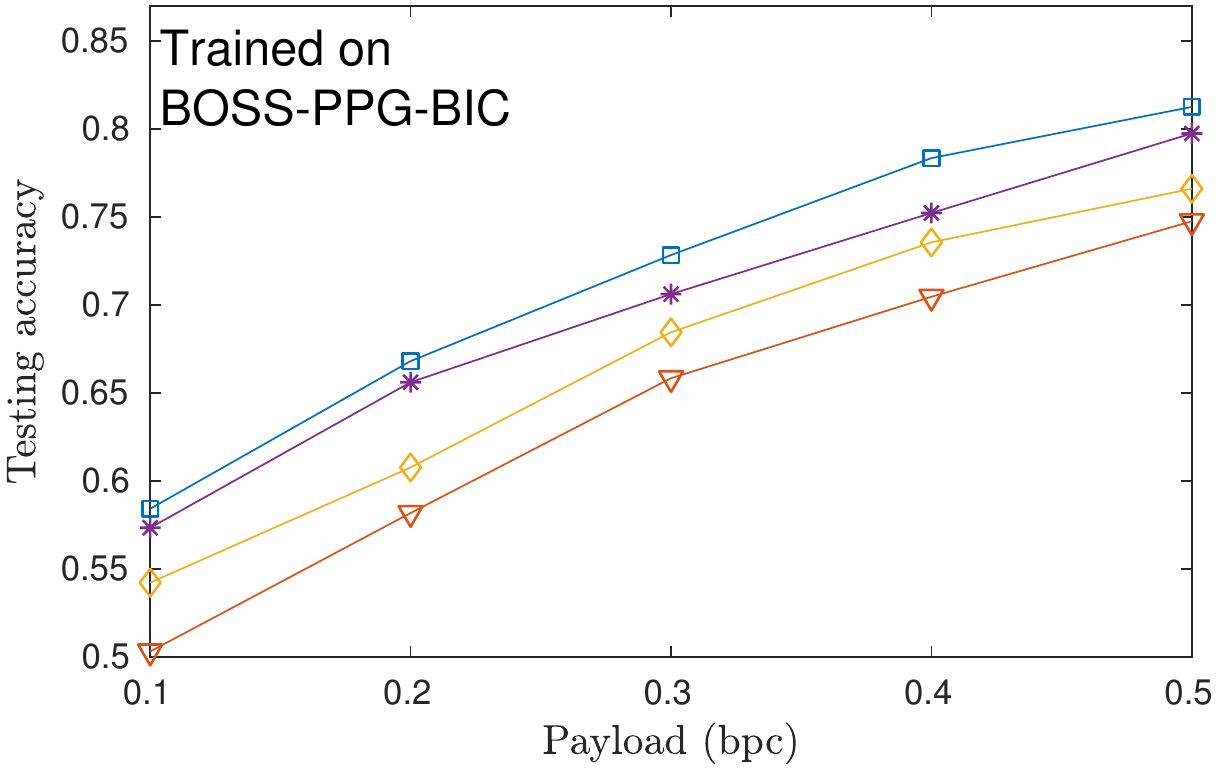}
  }\\
  \includegraphics[scale=0.8]{fig_full_comp_legend} 
  \caption[]{Comparison of performance of our proposed WISERNet with
    other state-of-the-art steganalyzers under cover-source
    mismatching scenario. The target is
    CMD-C-HILL. \subref{fig:full_comp_cover_source_mismatching_b}
    Trained on BOSS-AHD-LAN, while tested on SZUBASE-PPG-LAN;
    \subref{fig:full_comp_cover_source_mismatching_c} Trained on
    BOSS-PPG-BIC, while tested on SZUBASE-PPG-LAN.}
  \label{fig:full_comp_cover_source_mismatching}
\end{figure*}

We collected another 100,000 diverse raw images and followed the
dataset generating process as mentioned in
Sect.~\ref{sec:setups}~(with PPG demosaicking algorithm and
``Lanczos'' down-sampling kernel) to construct a new dataset
SZUBASE-PPG-LAN. Fig.~\ref{fig:full_comp_cover_source_mismatching} is
devoted to the comparison of performance of our proposed WISERNet with
other state-of-the-art steganalyzers on SZUBASE-PPG-LAN, under
cover-source mismatching scenarios. From
Fig.~\ref{fig:full_comp_cover_source_mismatching} we can see that the
impact of varing demosaicking algorithms is moderate, while the impact
of varing down-sampling kernels is more obvious. However, under such
cover-source mismatching scenarios, the impacts on the steganalyzers,
either hand-crafted or deep-learning based~(including our proposed
WISERNet), are similar. Our proposed WISERNet is still the one with
the best performance.
\section{Concluding remarks}
\label{sec:conclude}

Along with the arise of steganographic algorithms purposely for color
spatial images, the corresponding requirement for powerful color image
steganalysis becomes more compelling. In this paper we propose
WISERNet, the wider separate-then-reunion network for steganalysis of
color images. The major contributions of this work are as follows:
\begin{itemize}
\item We have provided theoretical rationale to claim that the
  summation in normal convolution actually impairs the signal-to-noise
  ratio, which collides with the main purpose of the bottom
  convolutional layer.
\item We have pointed out that the summation in normal convolution is a
  ``linear collusion attack'' which is a double-edged sword for color
  image steganalysis. Accordingly we have proposed WISERNet, the wider
  separate-then-reunion network for steganalysis of color images.
\item We have conducted extensive experiments on image datasets with
  different demosaicking and down-sampling opinions. The experimental
  results demonstrated the superiority of our proposed WISERNet.
\end{itemize}

Our future work will focus on two aspects: (1) making WISERNet capable
of identifying suspicious images under more complex cover source
mismatching scenarios; (2) further developing deep-learning
architectures suitable for large-scale JPEG color image steganalysis
on the basis of WISERNet.


\bibliographystyle{IEEEtran}
\bibliography{IEEEfull,csdl_final}

\begin{thebibliography}{10}
\providecommand{\url}[1]{#1}
\csname url@samestyle\endcsname
\providecommand{\newblock}{\relax}
\providecommand{\bibinfo}[2]{#2}
\providecommand{\BIBentrySTDinterwordspacing}{\spaceskip=0pt\relax}
\providecommand{\BIBentryALTinterwordstretchfactor}{4}
\providecommand{\BIBentryALTinterwordspacing}{\spaceskip=\fontdimen2\font plus
\BIBentryALTinterwordstretchfactor\fontdimen3\font minus
  \fontdimen4\font\relax}
\providecommand{\BIBforeignlanguage}[2]{{%
\expandafter\ifx\csname l@#1\endcsname\relax
\typeout{** WARNING: IEEEtran.bst: No hyphenation pattern has been}%
\typeout{** loaded for the language `#1'. Using the pattern for}%
\typeout{** the default language instead.}%
\else
\language=\csname l@#1\endcsname
\fi
#2}}
\providecommand{\BIBdecl}{\relax}
\BIBdecl
\renewcommand{\BIBentryALTinterwordstretchfactor}{4}

\bibitem{holub_eurasip_2014}
V.~Holub, J.~Fridrich, and T.~Denemark, ``Universal distortion function for
  steganography in an arbitrary domain,'' \emph{EURASIP Journal on Information
  Security}, vol. 2014, no.~1, pp. 1--13, 2014.

\bibitem{li_icip_2014}
B.~Li, M.~Wang, J.~Huang, and X.~Li, ``A new cost function for spatial image
  steganography,'' in \emph{Proc. IEEE 2014 International Conference on Image
  Processing, ({ICIP}'2014)}, 2014, pp. 4206--4210.

\bibitem{sedighi_tifs_2016}
V.~Sedighi, R.~Cogranne, and J.~Fridrich, ``Content-adaptive steganography by
  minimizing statistical detectability,'' \emph{{IEEE} Transactions on
  Information Forensics and Security}, vol.~11, no.~2, pp. 221--234, 2016.

\bibitem{fridrich_spie_2007}
J.~Fridrich and T.~Filler, ``Practical methods for minimizing embedding impact
  in steganography,'' in \emph{Proc. SPIE, Electronic Imaging, Security,
  Steganography, and Watermarking of Multimedia Contents IX}, vol. 6505, 2007,
  pp. 650\,502--1--650\,502--15.

\bibitem{denemark_ihmmsec_2015}
T.~Denemark and J.~Fridrich, ``Improving steganographic security by
  synchronizing the selection channel,'' in \emph{Proc. 3rd ACM Information
  Hiding and Multimedia Security Workshop (IH\&MMSec' 2015)}, 2015, pp. 5--14.

\bibitem{li_tifs_2015}
B.~Li, M.~Wang, X.~Li, S.~Tan, and J.~Huang, ``A strategy of clustering
  modification directions in spatial image steganography,'' \emph{{IEEE}
  Transactions on Information Forensics and Security}, vol.~10, no.~9, pp.
  1905--1917, 2015.

\bibitem{tang_spl_2016}
W.~Tang, B.~Li, W.~Luo, and J.~Huang, ``Clustering steganographic modification
  directions for color components,'' \emph{{IEEE} Signal Processing Letters},
  vol.~23, no.~2, pp. 197--201, 2016.

\bibitem{fridrich_tifs_2012}
J.~Fridrich and J.~Kodovsk\'y, ``Rich models for steganalysis of digital
  images,'' \emph{{IEEE} Transactions on Information Forensics and Security},
  vol.~7, no.~3, pp. 868--882, 2012.

\bibitem{holub_tifs_2013}
V.~Holub and J.~Fridrich, ``Random projections of residuals for digital image
  steganalysis,'' \emph{{IEEE} Transactions on Information Forensics and
  Security}, vol.~8, no.~12, pp. 1996--2006, 2013.

\bibitem{kodovsky_tifs_2012}
J.~Kodovsk\'y and J.~Fridrich, ``Ensemble classifiers for steganalysis of
  digital media,'' \emph{{IEEE} Transactions on Information Forensics and
  Security}, vol.~7, no.~2, pp. 432--444, 2012.

\bibitem{goljan_wifs_2014}
M.~Goljan, J.~Fridrich, and R.~Cogranne, ``Rich model for steganalysis of color
  images,'' in \emph{Proc. IEEE 2012 International Workshop on Information
  Forensic and Security ({WIFS}'2014)}, 2014, pp. 185--190.

\bibitem{goljan_spie_2015}
M.~Goljan and J.~J. Fridrich, ``{CFA}-aware features for steganalysis of color
  images,'' in \emph{Proc. IS\&T/SPIE Electronic Imaging 2015~(Media
  Watermarking, Security, and Forensics)}, 2015, pp. 94\,090V--1--94\,090V--13.

\bibitem{abdulrahman_iwcc_2015}
H.~Abdulrahman, M.~Chaumont, P.~Montesinos, and B.~Magnier, ``Color image
  stegananalysis using correlations between {RGB} channels,'' in \emph{Proc.
  IEEE 10th International Conference on Availability, Reliability and Security
  ({ARES}'2015)}, 2015, pp. 448--454.

\bibitem{abdulrahman_scn_2016}
------, ``Color images steganalysis using {RGB} channel geometric
  transformation measures,'' \emph{Security and communication networks},
  vol.~9, no.~15, pp. 2945--2956, 2016.

\bibitem{abdulrahman_ihmmsec_2016}
------, ``Color image steganalysis based on steerable {Gaussian} filters
  bank,'' in \emph{Proc. 4th ACM Information Hiding and Multimedia Security
  Workshop (IH\&MMSec'2016)}, 2016, pp. 109--114.

\bibitem{denemark_wifs_2014}
T.~Denemark, V.~Sedighi, V.~Holub, R.~Cogranne, and J.~Fridrich,
  ``Selection-channel-aware rich model for steganalysis of digital images,'' in
  \emph{Proc. 6th IEEE International Workshop on Information Forensic and
  Security ({WIFS}'2014)}, 2014, pp. 48--53.

\bibitem{tang_tifs_2016}
W.~Tang, H.~Li, W.~Luo, and J.~Huang, ``Adaptive steganalysis based on
  embedding probabilities of pixels,'' \emph{{IEEE} Transactions on Information
  Forensics and Security}, vol.~11, no.~4, pp. 734--745, 2016.

\bibitem{boroumand_tifs_2018}
M.~Boroumand and J.~Fridrich, ``Applications of explicit non-linear feature
  maps in steganalysis,'' \emph{{IEEE} Transactions on Information Forensics
  and Security}, vol.~13, no.~4, pp. 823--833, 2018.

\bibitem{schmidhuber_nn_2015}
J.~Schmidhuber, ``Deep learning in neural networks: An overview,'' \emph{Neural
  Networks}, vol.~61, pp. 85--117, 2015.

\bibitem{tan_apsipa_2014}
S.~Tan and B.~Li, ``Stacked convolutional auto-encoders for steganalysis of
  digital images,'' in \emph{Proc. Asia-Pacific Signal and Information
  Processing Association Annual Summit and Conference ({APSIPA}'2014)}, 2014,
  pp. 1--4.

\bibitem{qian_spie_2015}
Y.~Qian, J.~Dong, W.~Wang, and T.~Tan, ``Deep learning for steganalysis via
  convolutional neural networks,'' in \emph{Proc. IS\&T/SPIE Electronic Imaging
  2015~(Media Watermarking, Security, and Forensics)}, 2015, pp.
  94\,090J--1--94\,090J--10.

\bibitem{pibre_ei_2016_ver2}
L.~Pibre, P.~J\'{e}r\^{o}me, D.~Ienco, and M.~Chaumont, ``Deep learning is a
  good steganalysis tool when embedding key is reused for different images,
  even if there is a cover source-mismatch,'' in \emph{Proc. Media
  Watermarking, Security, and Forensics, Part of {IS\&T} International
  Symposium on Electronic Imaging (EI'2016)}, 2016, pp. 1--11.

\bibitem{qian_icip_2016}
Y.~Qian, J.~Dong, W.~Wang, and T.~Tan, ``Learning and transferring
  representations for image steganalysis using convolutional neural network,''
  in \emph{Proc. IEEE 2016 International Conference on Image Processing,
  ({ICIP}'2016)}, 2016, pp. 2752--2756.

\bibitem{xu_spl_2016}
G.~Xu, H.~Z. Wu, and Y.~Q. Shi, ``Structural design of convolutional neural
  networks for steganalysis,'' \emph{{IEEE} Signal Processing Letters},
  vol.~23, no.~5, pp. 708--712, 2016.

\bibitem{ye_tifs_2017}
J.~Ye, J.~Ni, and Y.~Yi, ``Deep learning hierarchical representations for image
  steganalysis,'' \emph{{IEEE} Transactions on Information Forensics and
  Security}, vol.~12, no.~11, pp. 2545--2557, 2017.

\bibitem{zeng_ei_2017}
J.~Zeng, S.~Tan, and B.~Li, ``Pre-training via fitting deep neural network to
  rich-model features extraction procedure and its effect on deep learning for
  steganalysis,'' in \emph{Proc. Media Watermarking, Security, and Forensics,
  Part of {IS\&T} International Symposium on Electronic Imaging (EI'2017)},
  2017, pp. 44--49.

\bibitem{zeng_tifs_2018}
J.~Zeng, S.~Tan, B.~Li, and J.~Huang, ``Large-scale {JPEG} steganalysis using
  hybrid deep-learning framework,'' \emph{{IEEE} Transactions on Information
  Forensics and Security}, vol.~13, no.~5, pp. 1242--1257, 2018.

\bibitem{chen_ihmmsec_2017}
M.~Chen, V.~Sedighi, M.~Boroumand, and J.~Fridrich, ``{JPEG}-phase-aware
  convolutional neural network for steganalysis of {JPEG} images,'' in
  \emph{Proc. 5th ACM Information Hiding and Multimedia Security Workshop
  (IH\&MMSec'2017)}, 2017, pp. 75--84.

\bibitem{he_cvpr_2016}
K.~He, X.~Zhang, S.~Ren, and J.~Sun, ``Deep residual learning for image
  recognition,'' in \emph{Proc. IEEE Conference on Computer Vision and Pattern
  Recognition (CVPR' 2016)}, 2016, pp. 770--778.

\bibitem{xu_ihmmsec_2017}
G.~Xu, ``Deep convolutional neural network to detect {J-UNIWARD},'' in
  \emph{Proc. 5th ACM Information Hiding and Multimedia Security Workshop
  (IH\&MMSec'2017)}, 2017, pp. 67--73.

\bibitem{su_tm_2005}
K.~Su, D.~Kundur, and D.~Hatzinakos, ``Statistical invisibility for
  collusion-resistant digital video watermarking,'' \emph{{IEEE} Transactions
  on Multimedia}, vol.~7, no.~1, pp. 43--51, 2005.

\bibitem{ioffe_icml_2015}
S.~Ioffe and C.~Szegedy, ``Batch normalization: Accelerating deep network
  training by reducing internal covariate shift,'' in \emph{Proc. International
  Conference on Machine Learning (ICML' 2015)}, 2015, pp. 448--456.

\bibitem{filler_tifs_2010}
T.~Filler and J.~Fridrich, ``Gibbs construction in steganography,''
  \emph{{IEEE} Transactions on Information Forensics and Security}, vol.~5,
  no.~4, pp. 705--720, 2010.

\bibitem{gonzalez_dip_2002}
R.~C. Gonzalez and R.~E. Woods, \emph{Digital Image Processing (3rd Ed)}.\hskip
  1em plus 0.5em minus 0.4em\relax Prentice Hall, New York, 2008, p. 354.

\bibitem{carey_conv_corr}
G.~Carey, ``Linear transformations and linear composites,''
  \url{http://psych.colorado.edu/~carey/Courses/PSYC7291/handouts/transformations.pdf},
  accessed: 2018-02-06.

\bibitem{pevny_tifs_2010}
T.~Pevn\'y, P.~Bas, and J.~Fridrich, ``Steganalysis by subtractive pixel
  adjacency matrix,'' \emph{{IEEE} Transactions on Information Forensics and
  Security}, vol.~5, no.~2, pp. 215--224, 2010.

\bibitem{pevny_ih_2008}
T.~Pevn\'y and J.~Fridrich, ``Benchmarking for steganography,'' in \emph{Proc.
  10th Information Hiding Workshop ({IH}'2008)}, 2008, pp. 251--267.

\bibitem{bas_ih_2011}
P.~Bas, T.~Filler, and T.~Pevn\'y, ``Break our steganographic system---the ins
  and outs of organizing {BOSS},'' in \emph{Proc. 13th Information Hiding
  Workshop ({IH}'2011)}, 2011, pp. 59--70.

\bibitem{jia_acmmm_2014}
Y.~Jia \emph{et~al.}, ``Caffe: Convolutional architecture for fast feature
  embedding,'' in \emph{Proc. 22nd ACM International Conference on Multimedia},
  2014, pp. 675--678.

\bibitem{wiki_cos_sim}
``Cosine similarity --- {W}ikipedia{,} the free encyclopedia,''
  \url{https://en.wikipedia.org/wiki/Cosine_similarity}, accessed: 2018-02-06.

\end{thebibliography}


\begin{IEEEbiography}[{\includegraphics[width=1in,height=1.25in,clip,keepaspectratio]{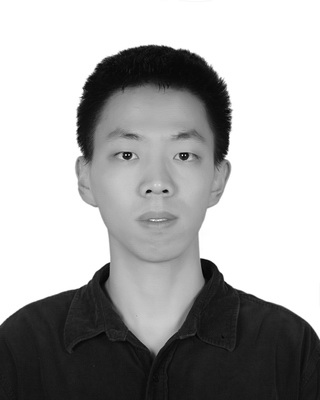}}]{Jishen Zeng (S'16)}
  received the B.S degree of electronic information science and
  technology from Sun Yat-sen University, Guangzhou, China in 2015. He
  is currently a Ph.D. student in Shenzhen University majoring in
  information and communication engineering. His current research
  interests include steganography, steganalysis, multimedia forensics,
  and deep learning.
\end{IEEEbiography}

\begin{IEEEbiography}[{\includegraphics[width=1in,height=1.25in,clip,keepaspectratio]{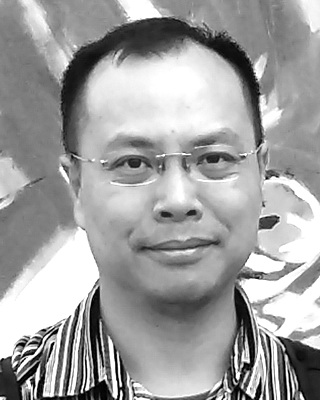}}]{Shunquan Tan (M'10--SM'17)}
  received the B.S. degree in computational mathematics and applied
  software and the Ph.D. degree in computer software and theory from
  Sun Yat-sen University, Guangzhou, China, in 2002 and 2007,
  respectively.

  He was a Visiting Scholar with New Jersey Institute of Technology,
  Newark, NJ, USA, from 2005 to 2006. He is currently an Associate
  Professor with College of Computer Science and Software Engineering,
  Shenzhen University, China, which he joined in 2007. His current
  research interests include multimedia security, multimedia
  forensics, and machine learning.
\end{IEEEbiography} 

\begin{IEEEbiography}[{\includegraphics[width=1in,height=1.25in,clip,keepaspectratio]{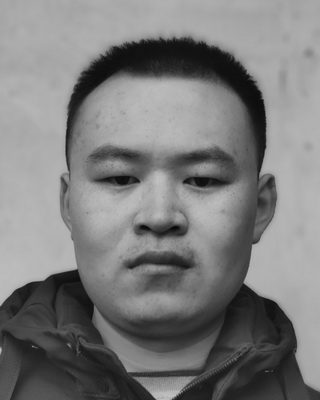}}]{Guangqing Liu}
  received the B.S degree of computer science and technology from
  Shenzhen University, Shenzhen, China in 2018. His current research
  interests include information hiding, multimedia forensics, and deep
  learning.
\end{IEEEbiography}

\begin{IEEEbiography}[{\includegraphics[width=1in,height=1.25in,clip,keepaspectratio]{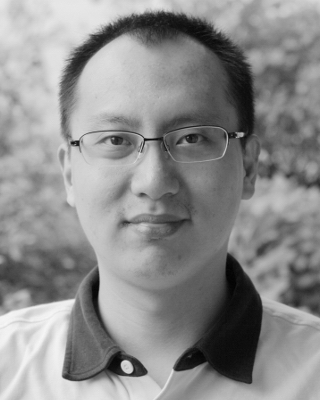}}]{Bin Li (S'07--M'09--SM'17)}
  received the B.E. degree in communication engineering and the
  Ph.D. degree in communication and information system from Sun
  Yat-sen University, Guangzhou, China, in 2004 and 2009,
  respectively. He was a Visiting Scholar with New Jersey Institute of
  Technology, Newark, NJ, USA, from 2007 to 2008. He is currently an
  Associate Professor with Shenzhen University, Shenzhen, China, where
  he joined in 2009. He is the director of Shenzhen Key Laboratory of
  Media Security. He is also a scholar with Peng Cheng Laboratory. He
  is now a member of IEEE Information Forensic and Security Technical
  Committee (IFS-TC).
 
  His current research interests include image processing, multimedia
  forensics, and pattern recognition.
\end{IEEEbiography}

\begin{IEEEbiography}[{\includegraphics[width=1in,height=1.25in,clip,keepaspectratio]{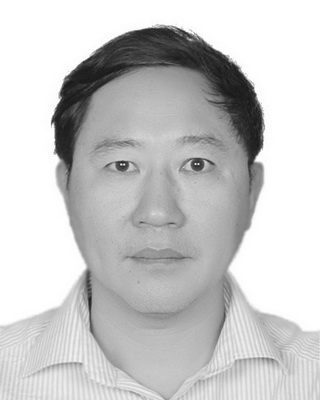}}]{Jiwu Huang (M'98--SM'00--F'16) }
  received the B.S. degree from Xidian University, Xian, China, in
  1982, the M.S. degree from Tsinghua University, Beijing, China, in
  1987, and the Ph.D. degree from the Institute of Automation, Chinese
  Academy of Sciences, Beijing, in 1998. He is currently a Professor
  with the College of Information Engineering, Shenzhen University,
  Shenzhen, China. His current research interests include multimedia
  forensics and security. He is a member IEEE Signal Processing
  Society Information Forensics and Security Technical Committee and
  serves as an Associate Editor for the IEEE Transactions on
  Information Forensics and Security. He was a General Co-Chair of the
  IEEE Workshop on Information Forensics and Security in 2013 and a
  TPC Co-Chair of the IEEE Workshop on Information Forensics and
  Security in 2018.
\end{IEEEbiography}

\vfill  
\end{document}